\documentclass[a4paper]{article}
\usepackage{graphicx}
\usepackage{times}
\usepackage{fullpage}

\usepackage{natbib}
\usepackage[superscript]{cite}
\usepackage{amsmath,amssymb}
\usepackage{caption}
\usepackage[usenames]{color}
\definecolor{gray}{rgb}{0.75,0.75,0.75}
\definecolor{blue}{rgb}{0,0,1}
\definecolor{blue2}{rgb}{0.68,0.84,0.9}
\definecolor{green}{rgb}{0.0,0.45,0.0}
\definecolor{green2}{rgb}{0.0,0.95,0.0}
\definecolor{red}{rgb}{1,0,0.0}
\definecolor{yellow}{rgb}{1,1,0.0}

\usepackage{soul}
\usepackage{lipsum}

\begin{document}

\begin{center}

\textbf{\Large {Modeling earthquakes with off-fault damage using the combined finite-discrete element method}} \\[20pt]
Kurama Okubo,$^{1}$ Esteban Rougier,$^2$, Zhou Lei, $^2$, Harsha S. Bhat$^{3}$
\end{center}
\vspace{-0.2cm}
\begin{enumerate}
\small
\it
\itemsep0em 
\item{Earth and Planetary Sciences, Harvard University, Cambridge, MA, USA}
\item{EES-17 - Earth and Environmental Sciences Division, Los Alamos National Laboratory, New Mexico, USA}
\item{Laboratoire de G\'{e}ologie, \'{E}cole Normale Sup\'{e}rieure, CNRS-UMR 8538, PSL Research University, Paris, France.}
\end{enumerate}

%\vspace{1cm}
%Last modified \today
\baselineskip14pt

%\clearpage

\noindent
% \hlg{TO DO: Harsha \& Faisal}~~~~\hly{TO DO: Faisal}~~~~\hlc{TO DO: Harsha}
%%%%%%%%%%%%%%%%%%%%%%%%%%%%%%%%%%%%%%%%%%%%%%%%%%%%%%%%%%%%%%%%%%%%%%
% INTRO PARAGRAPH (200-300 words)
%%%%%%%%%%%%%%%%%%%%%%%%%%%%%%%%%%%%%%%%%%%%%%%%%%%%%%%%%%%%%%%%%%%%%%
\noindent{\bf When a dynamic earthquake rupture propagates on a fault in the Earth's crust, the medium around the fault is dynamically damaged due to stress concentrations around the rupture tip. Recent field observations, laboratory experiments and canonical numerical models show the coseismic off-fault damage is essential to describe the coseismic off-fault deformation, rupture dynamics, radiation and overall energy budget. However, the numerical modeling of ``localized'' off-fault fractures remains a challenge mainly because of computational limitations and model formulation shortcomings. We thus developed a numerical framework for modeling coseismic off-fault fracture networks using the combined finite-discrete element method (FDEM) and we applied it to simulate dynamic ruptures with coseismic off-fault damage on various fault configurations. This paper addresses the role of coseismic off-fault damage on rupture dynamics associated with a planar fault, as a base case, and with a number of first-order geometrical complexities, such as fault kink, step-over and roughness.
}
%%%%%%%%%%%%%%%%%%%%%%%%%%%%%%%%%%%%%%%%%%%%%%%%%%%%%%%%%%%%%%%%%%%%%%
% MAIN TEXT (~ 4.5 pages when typeset)
%%%%%%%%%%%%%%%%%%%%%%%%%%%%%%%%%%%%%%%%%%%%%%%%%%%%%%%%%%%%%%%%%%%%%%
\vspace{0.2cm}
\noindent

%%%%%%%%%%%%%%%%%%%%%%%%%%%%%%%%%%%%%%%%%%%%%%%%%%%%%%%%%%%%%%%%%%%%%%%%%%%%%%%%%
%%%%%%%%%%%%%%%%%%%%%%%%%%%%%%%%%%%%%%%%%%%%%%%%%%%%%%%%%%%%%%%%%%%%%%%%%%%%%%%%%
\section{Introduction}
\label{intro}
%%%%%%%%%%%%%%%%%%%%%%%%%%%%%%%%%%%%%%%%%%%%%%%%%%%%%%%%%%%%%%%%%%%%%%%%%%%%%%%%%
%%%%%%%%%%%%%%%%%%%%%%%%%%%%%%%%%%%%%%%%%%%%%%%%%%%%%%%%%%%%%%%%%%%%%%%%%%%%%%%%%
\begin{figure*}
\center
\noindent\includegraphics[width=\textwidth]{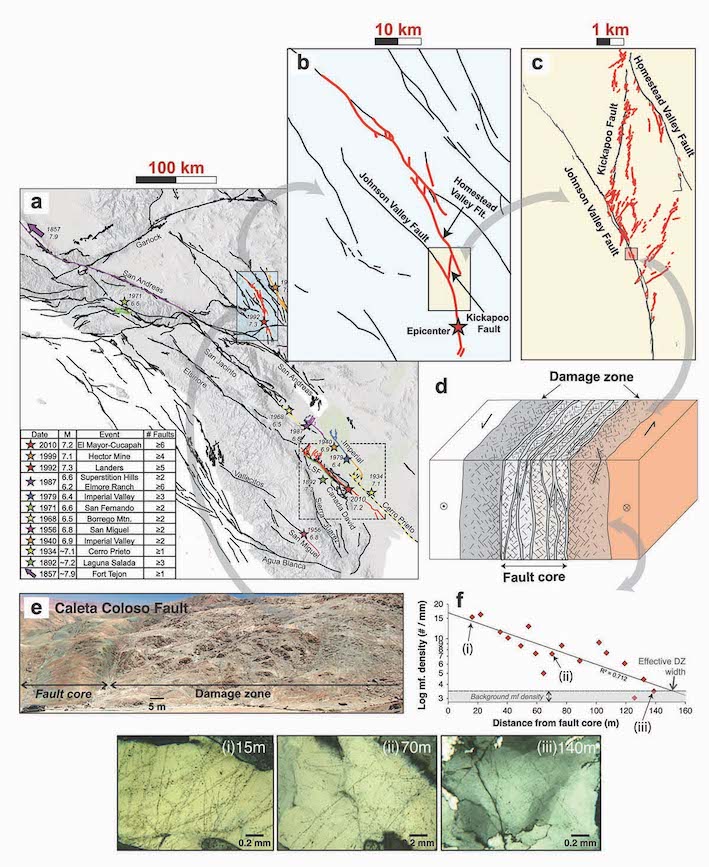}
\caption{Fault systems' hierarchical structure for a wide range of length scales. (a) Fault map of southern California \citep{fletcher2014}. Black lines indicate fault traces. Stars and color lines indicate the location of epicenters and rupture traces of historic earthquake events, respectively. (b) Fault map and rupture traces (in red) associated with the 1992 Landers earthquake \citep[modified from][]{sowers1994}. (c) Smaller scale off-fault fracture network  \citep{sowers1994}.  (d) Schematic of a typical fault zone structure, showing a fault core surrounded by damage zones \citep{mitchell2009}. (e, f) Fault damage zone of Caleta Coloso fault, the variation in microfracture (mf.) density within the damage zone as a function of distance from fault core, and some microfracture images taken at different distances from the fault core are shown.  \citep{mitchell2012a}.}
\label{fig:hierarchicalfaultsystem}
\end{figure*}
%%%%%%%%%%%%%%%%%%%%%%%%%%%%%%%%%%%%%%%%%%%%%%%%%%%%%%%%%%%%%%%%%%%%%%%%%%%%%%%%

The contribution of inelastic off-fault deformation to the rupture dynamics has been pointed out since 1970's. \citet{sibson1977} conceptually proposed a formulation for the overall energy budget of dynamic earthquake ruptures; a part of the energy released from accumulated strain energy by interseismic deformation is converted to seismic wave radiation, while the rest is expended in inelastic deformation processes within the fault zone. Then numerous studies via field observations have shown evidence of fractured rock surrounding the fault core, which can be coseismically damaged due to dynamic earthquake ruptures \citep[e.g.][]{chester1993, shipton2001, faulkner2011}. Furthermore, \citet{mitchell2009} showed that microfracture density is significantly higher in the near-fault region while it exponentially decreases with distance from the fault core, evidencing the presence of a well-defined off-fault damage zone.

Figure \ref{fig:hierarchicalfaultsystem} shows the hierarchical fault system in a wide range of length scales. Generally, fault geometrical complexity associated with an earthquake event is discussed in kilometric scale (Figures  \ref{fig:hierarchicalfaultsystem}a and  \ref{fig:hierarchicalfaultsystem}b). However, when we focus on a smaller portion of the fault system, we find off-fault fractures in subkilometric scale (Figures  \ref{fig:hierarchicalfaultsystem}c and \ref{fig:hierarchicalfaultsystem}d). These smaller scale off-fault fractures are either not included in kinematic and conventional dynamic earthquake rupture models, or their effects are homogenized using elastic-plastic constitutive damage models \citep[i.e.][]{andrews2005,templeton2008}. In these approaches the localized off-fault fractures indicated by red lines in Figure \ref{fig:hierarchicalfaultsystem}c remain to be fully modeled because of limitations in the damage model formulations, although their contributions might be significant on the rupture dynamics, deformation and residual stress field. Therefore, we need a numerical framework which allows for modeling both dynamic rupture on complex fault systems and coseismic generation of off-fault damage to investigate its effects on them.

This paper first describes the numerical framework of modeling dynamic earthquake rupture using the combined finite-discrete element method (FDEM) \citep{munjiza2004,munjiza2011,munjiza2015} that allows for modeling localized coseismic off-fault fractures and for quantifying their contributions to rupture dynamics, seismic radiation and energetics of earthquakes. We then perform a series of dynamic rupture modeling cases, showing FDEM's capability for modeling dynamic earthquake ruptures with dynamically activated tensile and shear fractures in the off-fault medium. Both friction and cohesion laws are applied on the fracture surfaces, providing a quantitative measure for energy dissipation due to the off-fault fracturing.

Since FDEM uses unstructured meshes, it can be used to model ruptures on complex fault systems such as fault kink, step-over and roughness. We demonstrate the rupture modeling on these fault configurations with first-order geometrical complexity in order to identify the damage pattern associated with each case and to investigate the rupture dynamics for first-order geometrical complexities. This case study eventually contributes to decompose the effects of coseismic off-fault damage on real fault systems with further parametric studies as a real fault system is formed by the aggregation of those simpler geometrical complexities.

%%%%%%%%%%%%%%%%%%%%%%%%%%%%%%%%%%%%%%%%%%%%%%%%%%%%%%%%%%%%%%%%%%%%%%%%%%%%%%%%
\section[Continuum-discontinuum approach]{Continuum-discontinuum approach for dynamic earthquake rupture modeling}
\label{sec:methodorogyintro}
%%%%%%%%%%%%%%%%%%%%%%%%%%%%%%%%%%%%%%%%%%%%%%%%%%%%%%%%%%%%%%%%%%%%%%%%%%%%%%%%
In the numerical framework of modeling both dynamic earthquake rupture and coseismic off-fault damage, geological faults and off-fault fractures are equivalently defined as discontinuities within a continuum medium. From this perspective, we consider both the faults and the off-fault damage in the same framework as an aggregation of fractures at different length scales. The activation of new fractures in the medium is represented as the loss of cohesive resistance. Frictional processes then take place at the boundary of the fractured surfaces, and they have a significant contribution in earthquakes' overall energy budget. Therefore, we need a modeling scheme able to handle both continuum (deformation) as well as discontinuum processes (fractures) within the same framework. Furthermore, this model requires efficient contact algorithms to resolve contact, cohesive and frictional forces, operating on every fracture surface and potential failure planes. We first provide a general description of the numerical framework using FDEM to model the dynamic earthquake ruptures on the prescribed fault system.

%%%%%%%%%%%%%%%%%%%%%%%%%%%%%%%%%%%%%%%%%%%%%%%%%%%%%%%%%%%%%%%%%%%%%%%%%%%%%%%%
\subsection{Formulation of FDEM}
%%%%%%%%%%%%%%%%%%%%%%%%%%%%%%%%%%%%%%%%%%%%%%%%%%%%%%%%%%%%%%%%%%%%%%%%%%%%%%%%
The application of FDEM, pioneered by \citet{munjiza1995}, has been expanded in the last couple of decades to solve broad scientific problems associated with fracturing and failure of solid media such as block caving, rock blasting, dam stability, rock slope stability and hydraulic fracturing \citep[e.g.][]{mahabadi2014, lisjak2014, lei2014, zhao2014,lei2018,lei2019,rougier2014,gao2019, euser2019}. In the FDEM framework, a solid medium is firstly discretized into finite elements, in which the deformation is governed by stress-strain constitutive laws as in the conventional finite element method (FEM).
The interaction among individual elements is then computed based of prescribed cohesion and friction laws.
In this study, we utilized the FDEM-based software tool, HOSSedu (Hybrid Optimization Software Suite - Educational Version), developed by Los Alamos National Laboratory (LANL) \citep{hoss2015}. More details of main algorithmic solutions used within HOSSedu can be found in a series of monographs \citep{munjiza2004, munjiza2011, munjiza2015}.

%%%%%%%%%%%%%%%%%%%%%%%%%%%%%%%%%%%%%%%%%%%%%%%%%%%%%%%%%%%%%%%%%%%%%%%%%%%%%%%%
\subsection{Model description}
\begin{figure}
\center
\noindent\includegraphics[width=0.5\textwidth]{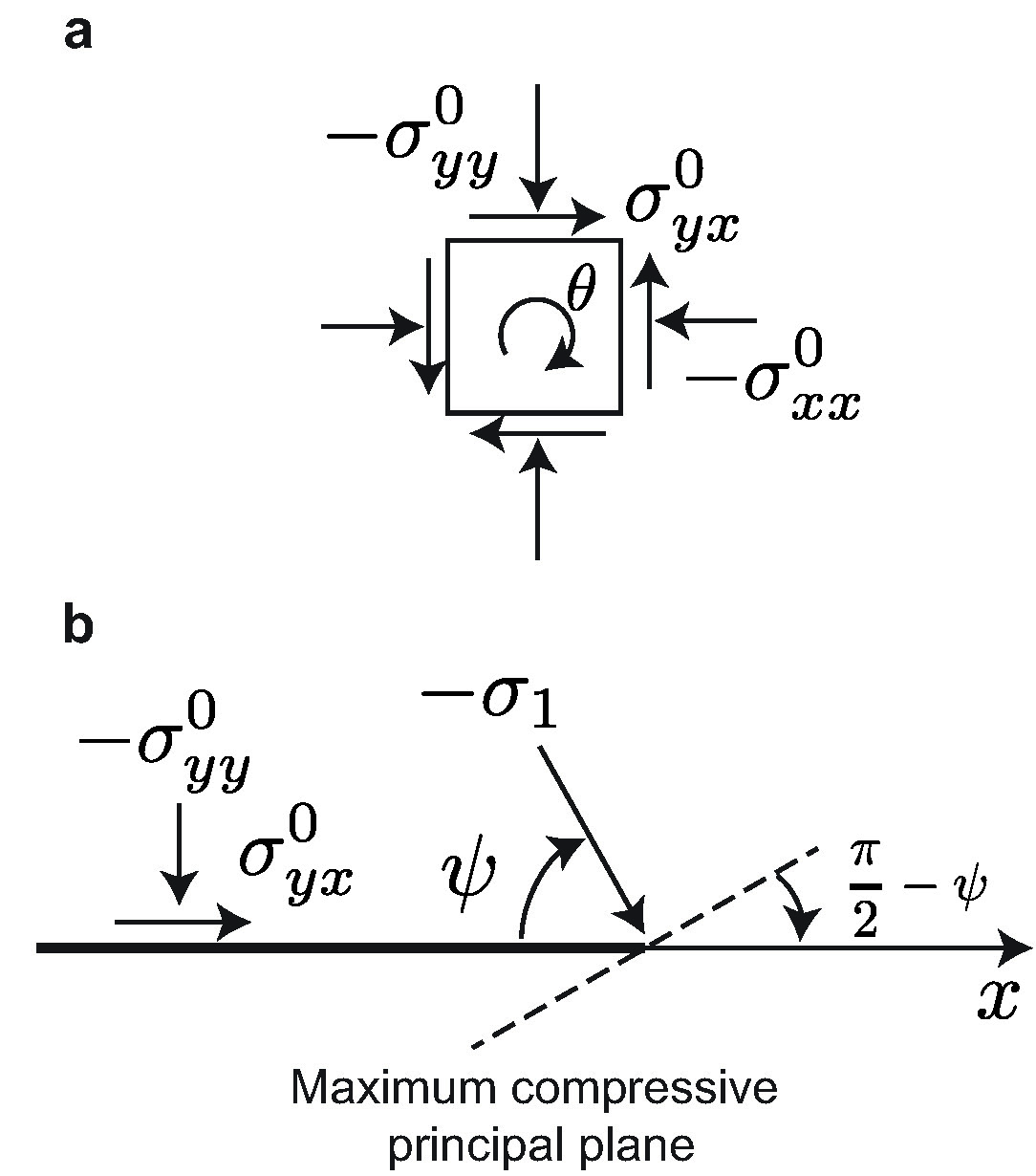}
\caption{Sign convention for stress and orientation. (a) Tensile and clockwise directions are positive for stresses. (b) Sign convention for the stresses on the fault. $-\sigma_{yy}^0$ and $\sigma_{yx}^0$ are respectively initial normal traction and shear traction applied on the fault along the x axis. $-\sigma_1$ is maximum compressive principal stress with the angle $\psi$ to the fault.}
\label{fig:signconvention}
\end{figure}

%%%%%%%%%%%%%%%%%%%%%%%%%%%%%%%%%%%%%%%%%%%%%%%%%%%%%%%%%%%%%%%%%%%%%%%%%%%%%%%%
In this section we describe the prestress condition and failure criteria used for dynamic earthquake rupture modeling with coseismic off-fault damage. The sign convention used in this work considers that tensile stresses and clockwise rotations are positive as shown in Figure \ref{fig:signconvention}.

%%%%%%%%%%%%%%%%%%%%%%%%%%%%%%%%%%%%%%%%%%%%%%%%%%%%%%%%%%%%%%%%%%%%%%%%%%%%%%%%
\begin{figure}
\center
\noindent\includegraphics[width=\textwidth]{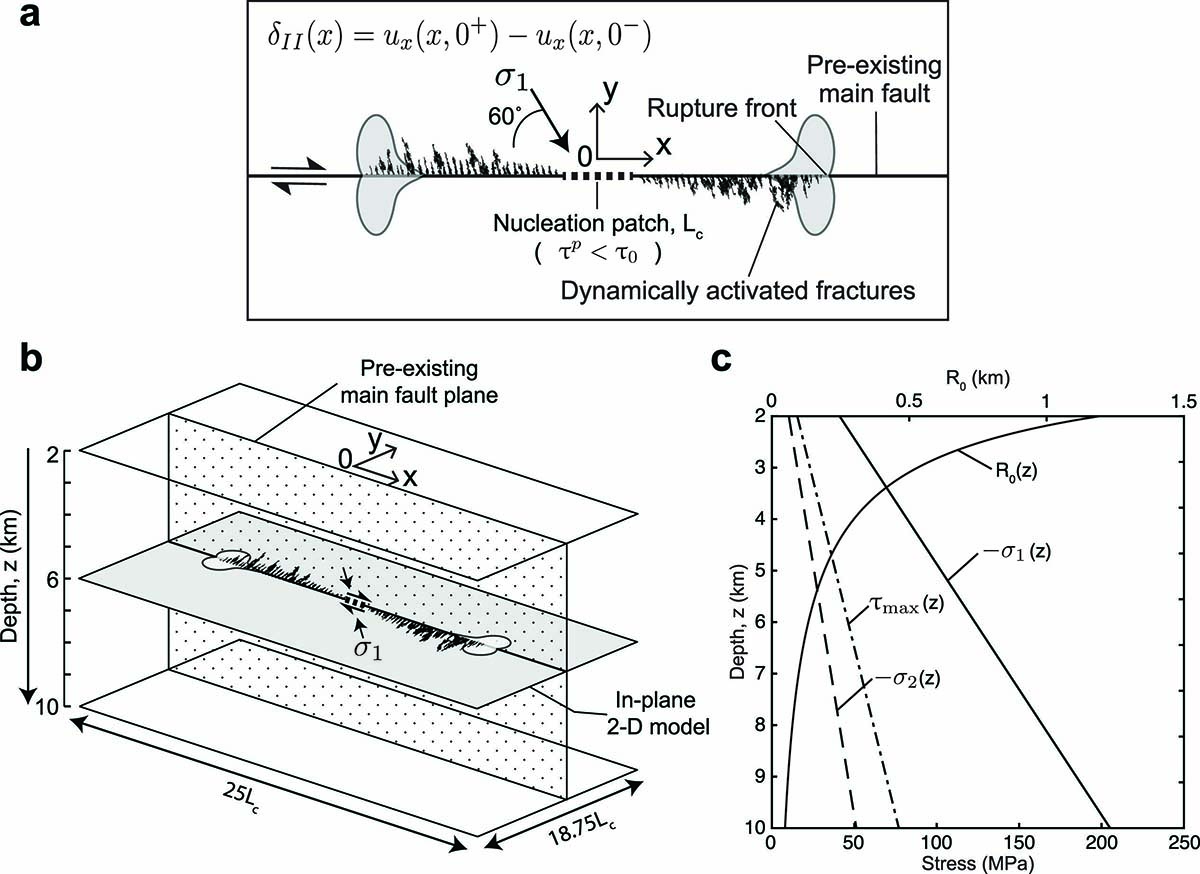}
\caption{Model description for the case study with depth. (a) 2-D strike-slip fault for dynamic rupture modeling with coseismic off-fault damage. The pre-existing fault is defined as the interface without cohesion. The orientation of maximum compressional principal stress $\sigma_1$ is fixed to 60$^\circ$ from the main fault. The slip on the fault $\delta_{II}$ is defined as the relative displacement.  (b) Schematic of case study with depth. Lc indicates the critical nucleation length at instability in equation \ref{eq:nuc}.  (c) The evolution of initial stress state and quasi-static process zone size $R_0(z)$ with depth. $-\sigma_1(z)$, $-\sigma_2(z)$, $\tau_{\text{max}}(z)$ indicate maximum principal stress, minimum principal stress and maximum shear traction, respectively.}
\label{fig:modeldescription}
\end{figure} 
%%%%%%%%%%%%%%%%%%%%%%%%%%%%%%%%%%%%%%%%%%%%%%%%%%%%%%%%%%%%%%%%%%%%%%%%%%%%%%%%
\subsection{Initial stress state with depth}
\label{sec:modeldescription}
%%%%%%%%%%%%%%%%%%%%%%%%%%%%%%%%%%%%%%%%%%%%%%%%%%%%%%%%%%%%%%%%%%%%%%%%%%%%%%%%
We follow a similar process to that proposed by \citet{templeton2008} and \citet{xu2012a} to make an assumption of initial stress state as a function of depth. For the sake of simplicity, we assume the prestress state linearly increases in depth based on lithostatic and hydrostatic conditions; therefore, it does not represent to a certain regional stress at depth.

A main fault plane is set parallel to the depth direction $z$ while the $xy$ - plane is perpendicular to $z$. The x-axis is aligned with the main fault and the origin of the x-y coordinate system is located in the middle of the main fault. The initial stress state is set for triggering a right-lateral strike-slip on the main fault. We solve this problem assuming plane strain conditions. The initial stress state is initially uniform in the homogeneous and isotropic elastic medium, and is given by
\begin{equation} \sigma_{ij}^0 = 
\left[
    \begin{array}{cc}
      \sigma_{xx}^0 & \sigma_{yx}^0 \\
      \sigma_{yx}^0 & \sigma_{yy}^0 
    \end{array}
  \right].
\end{equation}
Let normal stress $\sigma_{yy}^0$ on the main fault be given by linear overburden effective stress gradient such that 
\begin{equation}
\sigma _{yy}^0 = -(\rho - \rho_{w}) gz,
\label{eq:syy}
\end{equation}
where $\rho$ is the density of rock, $\rho_{w}$ is the density of water, $g$ is the gravitational acceleration and $z$ is the depth measured from the ground surface. 
The initial shear stress $\sigma_{yx}^0$ is estimated in terms of the seismic $S$ ratio, defined by \cite{andrews1976}, on the main fault such as
\begin{equation}
S = \frac{f_s(-\sigma_{yy}^0) - \sigma_{yx}^0}{\sigma_{yx}^0 - f_d(-\sigma_{yy}^0)},
\label{eq:Sratio}
\end{equation}
where $f_s$ and $f_d$ are the static and dynamic friction coefficients respectively. The value of the $S$ ratio defines whether the rupture velocity is supershear ($S < 1.77$), or remains sub-Rayleigh ($S > 1.77$) in 2-D.
Thus the initial shear stress on the main fault can be written as
\begin{equation}
\sigma_{yx}^0 = \frac{f_s + S f_d}{1+S} (-\sigma_{yy}^0).
\end{equation}

The horizontal compressive stress $\sigma_{xx}^0$ is then determined by the normal stress $\sigma_{yy}^0$, shear stress $\sigma_{yx}^0$  and the given orientation of the initial compressive principal stress to the main fault $\psi$ (indicated in Figure \ref{fig:signconvention}b) as follows:
\begin{equation}
\sigma_{xx}^0 = \left( 1 - \frac{2\sigma_{yx}^0}{\tan{(2\psi)} \sigma_{yy}^0} \right ) \sigma_{yy}^0.
\end{equation}
The relationship of the magnitude of $\sigma_{xx}^0$ and $\sigma_{yy}^0$ depends on $\psi$ in the following manner:
\begin{equation}
\left\{ \begin{array}{ll}
    (-\sigma_{xx}^0) \geq (-\sigma_{yy}^0), & \quad 0<\psi \leq \pi/4 \\
    \\
    (-\sigma_{xx}^0) < (-\sigma_{yy}^0), & \quad \pi/4 <\psi < \pi / 2 \\
  \end{array} \right.
\end{equation}
which is consistent with the condition of initial stress state defined by \citet{poliakov2002} and \citet{rice2005}.
%%%%%%%%%%%%%%%%%%%%%%%%%%%%%%%%%%%%%%%%%%%%%%%%%%%%%%%%%%%%%%%%%%%%%%%%%%%%%%%%
\subsection{Failure criteria}
\label{sec:failurecriteria}
%%%%%%%%%%%%%%%%%%%%%%%%%%%%%%%%%%%%%%%%%%%%%%%%%%%%%%%%%%%%%%%%%%%%%%%%%%%%%%%%
In the FDEM framework, cracks are represented as a loss of cohesion at the interfaces of the finite elements in the model. The combined single and smeared discrete crack approach \citep{munjiza1999} is generally accepted as a crack model based on fracture energy, where the cohesion and friction are prescribed following actual representations of experimental stress-strain curves \citep{lei2014}. 
It is worth noting that the cohesion and the friction against the opening or sliding motion between contactor and target are a function of displacements defined by the aperture $\delta_I$ and the slip $\delta_{II}$ between the contactor and the target.

\begin{figure}
\center
\noindent\includegraphics[width=\textwidth]{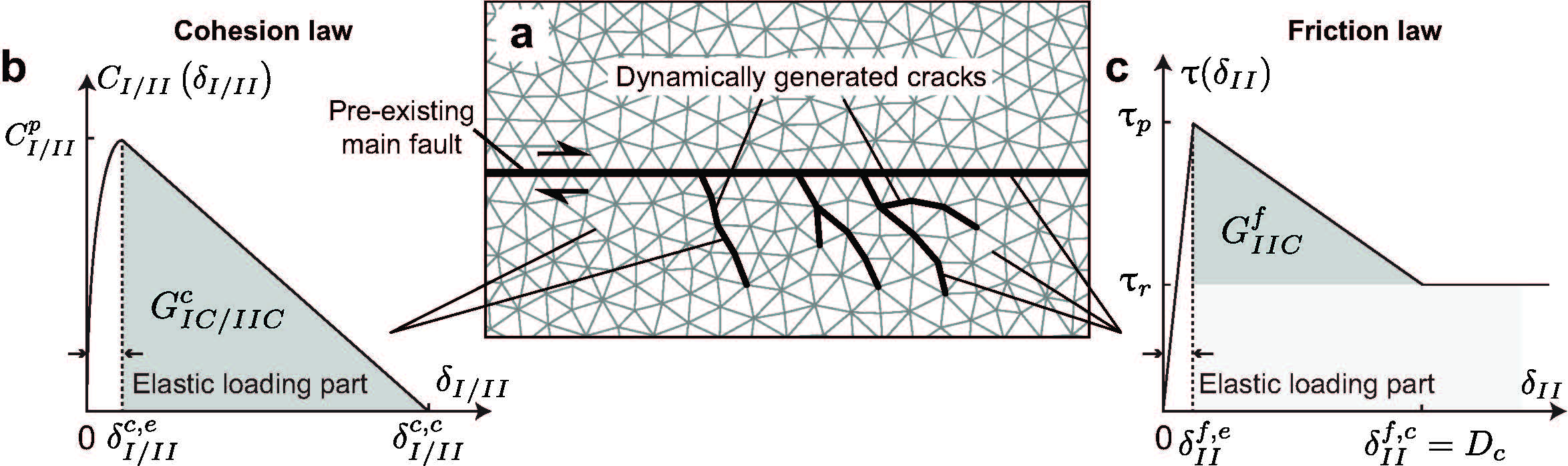}
\caption{Schematic of contact algorithm. (a) Computational domain discretized using an unstructured mesh. Every interface between elements is  regarded as a potential failure plane, where cohesion and friction stresses are operating as a function of displacements $\delta_{I/II}$. (b) Linear displacement softening cohesion law. The area highlighted in gray under the softening part of the curve indicates the fracture energy associated with cohesion in tension $G_{IC}^c$ and in shear $G_{IIC}^c$ respectively. (c) Linear slip-weakening law. The energy dissipated by frictional process is divided into the fracture energy associated with friction, $G_{IIC}^f$, while the rest is considered as heat.}
\label{fig:contactalgorithm}
\end{figure}

The cohesive and frictional resistances are applied on every interface between elements (i.e., at every edge), which is regarded as a potential failure plane. Both cohesion and friction curves are divided into two parts, an elastic loading part and a displacement-weakening part as shown in Figure \ref{fig:contactalgorithm}. In the elastic loading part, the resistant forces against displacements acting on the interface increase non-linearly (for the case of cohesion) or linearly (for the case of friction) with the stiffness of the elastic loading portions being $p^c, p^f$ respectively. Since this elastic loading part ideally should be zero to represent the material continuity, the stiffnesses, $p^c$ and $p^f$, are chosen to be much higher than the Young's modulus of the material $E$ in order to minimize the extra compliance introduced by the interfaces' elastic loading portions. In this study, we chose $p^c = 1000E$, and $p^f$ is chosen in the same order of $p^c$ as described in the following section. When the applied traction on the interface reaches the peak tensile or shear cohesion strengths $C_{I/II}^p$, the interface bonding starts to be weakened (i.e., damage starts to accumulate), and eventually it loses the cohesion (Figure \ref{fig:contactalgorithm}b). When the shear traction reaches to frictional strength $\tau_p$, it decreases down to the residual strength at critical displacements $D_c$ as shown in Figure \ref{fig:contactalgorithm}c. The friction curve follows the linear slip-weakening law originally proposed by \citet{ida1972a} and \citet{palmer1973}, which has been widely used for dynamic earthquake rupture modeling \citep[e.g.][]{andrews1976, aochi2002a, delapuente2009}. Eventually, the medium's shear strength follows the Mohr-Coulomb failure criteria. Note that the friction law is operating both on the main fault and the secondary cracks activated in the off-fault medium.

The mixed mode fracture is evaluated by a damage parameter, $D$, which is defined as
\begin{equation}
D_i = \frac{\delta_i - \delta_i^{c,e}}{\delta_i^{c,c} - \delta_i^{c,e}} \quad i = I,II
\label{eq:damagedef1}
\end{equation}
\begin{equation}
D = \sqrt{D_I^2 + D_{II}^2} ~(0 \leq D \leq 1)
\label{eq:damagedef2}
\end{equation}
\begin{equation}
D^T = \frac{D_I}{D} = \left\{\begin{array}{ll}
        1, & \text{for purely tensile crack}\\
        0, & \text{for purely shear crack}
                \end{array}\right\},
\label{eq:damagetype}
\end{equation}
where $D_i$ ($i = I, II$) are the components of damage for tensile and shear modes,  $\delta_i$ are the normal and tangential displacements, $\delta_i^{c, e}$ are the initial critical displacements for elastic loading, $\delta_{i}^{c,c} - \delta_{i}^{c,e}$ are the maximum displacements during linear-softening where $\delta_{i}^{c,c}$ are the initial critical displacements for the linear-weakening part,  $D$ is the degree of damage and $D^T$ indicates the type of damage.
Similar expressions can be found in \citet{rougier2011} and \citet{lisjak2014}.

Since we employed a linear softening law, the fracture energies related with cohesion for tensile (mode I) and shear (mode II) (i.e., the energy required to completely break the bonds between finite elements) are evaluated as
\begin{equation}
G_{iC}^c = \frac{1}{2} C^p_{i} \left ( \delta_{i}^{c,e} - \delta_{i}^{c,c} \right )  \quad i = I,II
\label{eq:GIC}
\end{equation}
where $G_{iC}^c$ are the tensile and the shear fracture energies and  $C^p_{i}$ are the tensile and the shear cohesive strengths.
The fracture energy for friction is, following \citet{palmer1973}, described as
\begin{equation}
G_{IIC}^f = \dfrac{1}{2} D_c \left(\tau_p - \tau_r \right)
\label{eq:GIICf}
\end{equation}
where $G_{IIC}^f$ is the fracture energy for friction, $D_c = \delta_{II}^{f, c}$ is the critical slip distance for friction and $\tau_p$ and $\tau_r$ are the peak strength and the residual strength for friction, defined as 
\begin{equation}
\tau_p = f_s (-\sigma_n)
\label{eq:peakfric}
\end{equation}
\begin{equation}
\tau_r = f_d (-\sigma_n),
\label{eq:residualfric}
\end{equation}
where $f_s$ and $f_d$ are the static and dynamic friction coefficients and $\sigma_n$ is the normal stress on the contact surface. Note that the elastic loading part $\delta_{i}^{f,e}$ is much smaller than $D_c$, so that the representation of fracture energy $G_{IIC}^{f}$ by equation (\ref{eq:GIICf}) is acceptable even without the consideration of elastic loading part.
%%%%%%%%%%%%%%%%%%%%%%%%%%%%%%%%%%%%%%%%%%%%%%%%%%%%%%%%%%%%%%%%%%%%%%%%%%%%%%%%
\subsection{Friction law}
%%%%%%%%%%%%%%%%%%%%%%%%%%%%%%%%%%%%%%%%%%%%%%%%%%%%%%%%%%%%%%%%%%%%%%%%%%%%%%%%
When the amount of slip exceeds the elastic slip distance for cohesion $\delta_{II}^{c,e}$, the cohesive force starts weakening. We assume that the friction starts weakening at $\delta_{II}^{f,e} = \delta_{II}^{c,e}$ so that the cohesion and the friction start weakening at the same amount of slip. We do this by adjusting the stiffness of elastic loading for friction $p^f$, as follows
\begin{equation}
p^f = \dfrac{\tau_{II}^{p}}{2C_{II}^{p}} p^c.
\label{eq:penalty_f}
\end{equation}
The fracture energy related with friction, $G_{IIC}^f$, is approximated from the equation (\ref{eq:GIICf}).

One interesting question is, as pointed out by \citet{rice2005}, what parameters vary with depth. In our parametrization, normal stress on the fault lithostatically increases with depth. \citet{lachenbruch1980a} proposed a formula of frictional resistance similar to the exponential slip-weakening law, where the slip-weakening distance $D_c$ on the fault is almost independent of depth because it is composed by physical parameters like the width of fault gouge and other coefficients related with pore fluid or rock material, which are assumed to be constant with depth (also referred in \citet{rice2005}). In this case, $G_{IIC}^f$ on the fault derived by equation (\ref{eq:GIICf}) increases with depth as the strength drop linearly increases as $\tau_p - \tau_r = (f_s - f_d) \left\{ -\sigma^0_{yy}(z) \right\} $ in our model description, described as
\begin{equation}
G_{IIC}^f(z) = \frac{1}{2}D_c^* \left(\tau_p - \tau_r \right),
\label{eq:GIICfindepth}
\end{equation}
where $D_c^*$ is a given constant critical slip distance with depth. 
This is the first scenario that we consider. The second scenario is to assume that $G_{IIC}^f$ on the fault is kept constant with depth. In this case, $D_c$ decreases with depth, as a function of a given constant $G_{IIC}^{f*}$ on the fault, as follows
\begin{equation}
D_c(z)= \frac{2G_{IIC}^{f*}}{\left( f_{s} - f_{d} \right) \left\{-\sigma^0_{yy}(z) \right\} }.
\label{eq:Dcwithdepth}
\end{equation}
For the sake of simplicity, we call the first scenario constant $D_c$ case, and the second scenario constant $G_{IIC}$ case.

In both scenarios, as proposed by \citet{palmer1973}, the process zone size $R_{0}$ for the quasi-stationary crack, over which the friction is weakened with ongoing slip to the residual strength, is described as
\begin{equation}
R_{0}(z) =  \frac{9\pi}{32(1-\nu)} \frac{\mu D_c^*}{(f_{s} - f_{d}) \left\{ -\sigma^0_{yy}(z) \right\} },
\label{eq:processzonesizeconstDc} 
\end{equation}
for the constant $D_c$ case, while
\begin{equation}
R_{0}(z) = \frac{9\pi}{16(1-\nu)} \frac{\mu G_{IIC,f}^*}{ \left[ (f_{s} - f_{d}) \left\{-\sigma^0_{yy}(z)\right\} \right] ^2},
\label{eq:processzonesizeconstGIIC}
\end{equation}
for the constant $G_{IIC}$ case. As shown by equations (\ref{eq:processzonesizeconstDc}) and (\ref{eq:processzonesizeconstGIIC}), $R_0$ decreases with depth as $\left\{ -\sigma^0_{yy}(z) \right\} ^{-1}$ for constant $D_c$ case and $\left\{ -\sigma^0_{yy}(z) \right\} ^{-2}$ for constant $G_{IIC}$ case.

Note that since the size of potential failure area is of the same order of magnitude as $R_{0}(z)$ \citep[e.g.][]{poliakov2002}, the damage zone is expected to decrease with depth, as mentioned by \citet{rice2005}. 

To artificially nucleate the rupture from a part of pre-exisisting fault, a slippery zone where frictional resistance is lower than the rest of the fault is set in the nucleation patch. The length of the slippery zone is slightly greater than the critical nucleation length at instability, $L_c$, derived by \citet{palmer1973} such as
\begin{equation}
L_c = \frac{2 \mu D_c (\tau_p-\tau_r)}{\pi (\sigma^0_{yx} - \tau_r)^2}.
\label{eq:nuc}
\end{equation}

%%%%%%%%%%%%%%%%%%%%%%%%%%%%%%%%%%%%%%%%%%%%%%%%%%%%%%%%%%%%%%%%%%%%%%%%%%%%%%%%
\subsection{Closeness to failure}
\begin{figure}
\center
\noindent\includegraphics[width=0.5\textwidth]{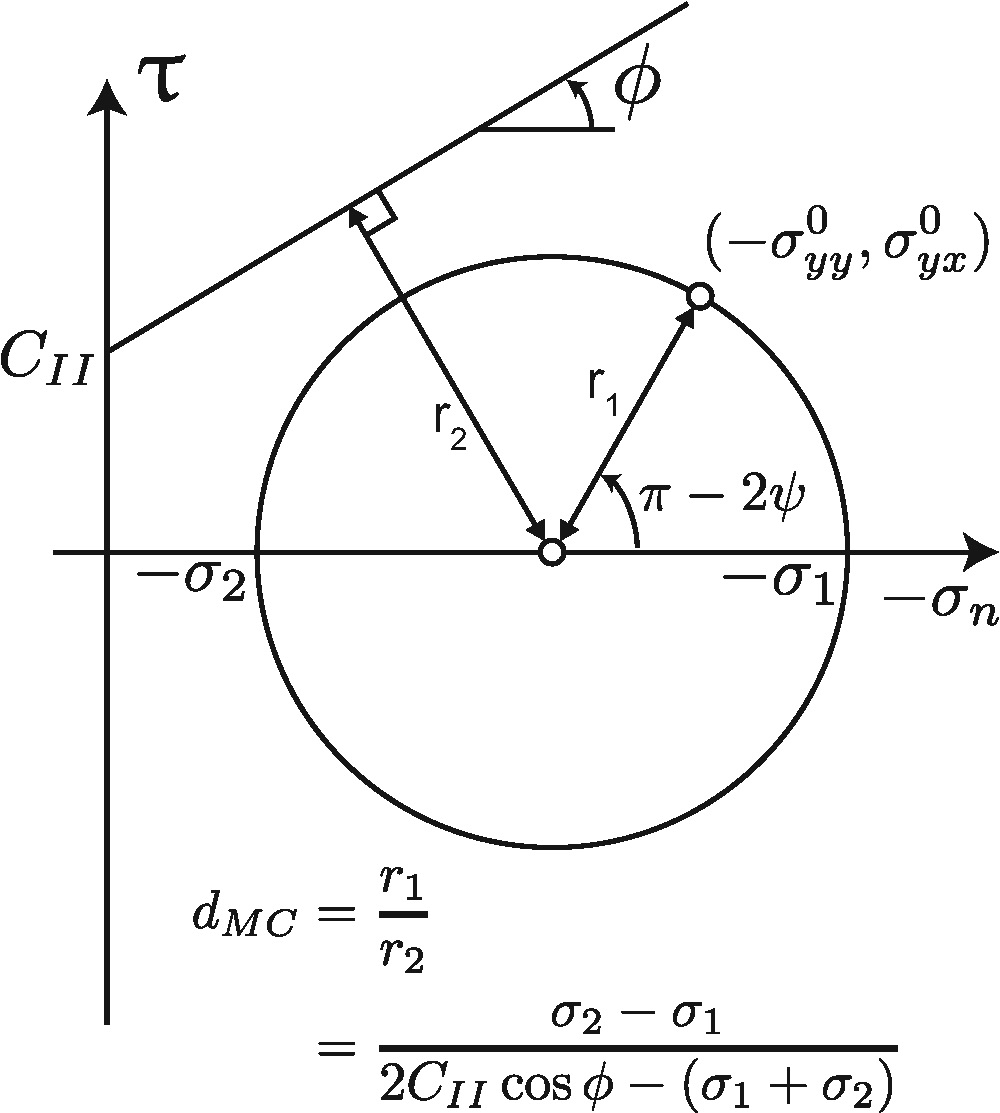}
\caption[Mohr-Coulomb failure criteria and closeness to failure.]{Mohr-Coulomb failure criteria and closeness to failure, $d_{MC}$.}
\label{fig:dMC}
\end{figure}

Here, we describe the parametrization of the failure criteria based on the fracture energy estimated from the experiments and observations \citep{viesca2015, passelegue2016b}, and the closeness to failure, proposed by \citet{templeton2008} and \citet{viesca2008}, which indicates the safety of the initial stress state to the failure of the material represented by the ratio of the radius of the Mohr's circle to the distance to the Mohr-Coulomb criteria.
Let $\sigma_1$ and $\sigma_2$ be the maximum and minimum compressive principal stresses. Assume a Mohr-Coulomb friction criteria with shear peak strength $C^p_{II}$. Then the closeness to failure, $d_{MC}$, is derived from geometrical relationships such that
\begin{align}\nonumber
d_{MC} &= \frac{\sigma_2 - \sigma_1}{2C^p_{II} \cos{\phi} - (\sigma_1 + \sigma_2)} \\
&= \dfrac{\left( \dfrac{\sigma_1}{\sigma_2} - 1 \right)}  {\left( \dfrac{\sigma_1}{\sigma_2} + 1 \right) - 2\left(\dfrac{C^p_{II}}{\sigma_2} \cos{\phi} \right)}
\label{eq:dMC}
\end{align}
where $\phi$ is the friction angle as $\tan{\phi} = f_{s}$ (Figure \ref{fig:dMC}). Thus $d_{MC}<1$ means no failure and $d_{MC}\ge1$ implies the initiation of failure in shear on the corresponding plane. Note that $d_{MC}$ locally changes due to perturbations of the stress field.

To make the medium equally close to failure, regardless of the stress state, $d_{MC}$ is kept constant with depth. By assuming the constant angle of maximum compressive principal stress $\Psi$ and the seismic ratio $S$, the ratio of principal stresses ${\sigma_{1}}/{\sigma_{2}}$ is derived to be constant with depth. Thus from equation (\ref{eq:dMC}), the ratio $C^p_{II}/\sigma_2 $ has to be kept constant to obtain an equal closeness to failure with depth, implying that peak cohesion $C^p_{II}$ must increase linearly in depth. Therefore we first calculate $\sigma_{ij}^0$ as described in previous section, and then we then derive $C^p_{II}$ as follows 
\begin{equation}
C^p_{II} = \frac{\sigma_2 - \sigma_1 + d_{MC} (\sigma_1 + \sigma_2) \sin{\phi}}{2 d_{MC} \cos{\phi}},
\label{eq:CFII}
\end{equation}
where $d_{MC}$ should be chosen carefully to avoid $C^p_{II}$ being negative. $C^p_{I}$ is chosen from the experiments \citep{cho2003}, and is kept constant with depth. See \citet{okubo2018h} for the rest of parameters to define cohesion curve such as $\delta_{I/II}^{c, e}$ and $\delta_{I/II}^{c, c}$.

%%%%%%%%%%%%%%%%%%%%%%%%%%%%%%%%%%%%%%%%%%%%%%%%%%%%%%%%%%%%%%%%%%%%%%%%%%%%%%%%
\section{Results}
\label{dyneq}
In this section we present the results obtained from FDEM simulations for a number of relevant earthquake rupture cases: 1) Rupture of a planar fault (base case), 2) Rupture along a fault with a kink, 3) Rupture on a step-over fault system, and 4) Rupture along a rough fault.
%%%%%%%%%%%%%%%%%%%%%%%%%%%%%%%%%%%%%%%%%%%%%%%%%%%%%%%%%%%%%%%%%%%%%%%%%%%%%%%%
\subsection{Rupture on a planar fault}
We performed the dynamic earthquake rupture modeling with a planar strike-slip fault, surrounded by intact rock, in plane strain conditions allowing for the generation of off-fault fractures. Figure \ref{fig:modeldescription}a shows the model description for the 2-D dynamic earthquake rupture modeling. In this paper, we show the result of 2km depth for sub-Rayleigh and supershear cases, where the height and width of model are 25Lc ($\sim$77km) and 18.75Lc ($\sim$58km) with sub-Rayleigh case, respectively. The material properties and parameters for contact interactions used in this section are listed in Tables \ref{tab:planarfaultparameters_material} and  \ref{tab:planarfaultparameters_contact}.

%%%%%%%%%%%%%%%%%%%%%%%%%%%%%%%%%%%%%%%%%%%%%%%%%%%%%%%%%%%%%%%%%%%%%%%%%%%%%%%%
%-------------------------------------------------------------------------%
\begin{table}
\center
\caption{Material constants and parameters for modeling a planar fault at 2km depth.}
\begin{tabular}{p{2.0cm} p{3cm} p{7.5cm}}
\hline
\textbf{Valuables} & \textbf{Values} & \textbf{Descriptions} \\ \hline
$E$ & 75 GPa & Young's modulus\\
$\mu$ & 30 GPa & Shear modulus\\
$\nu$ & 0.25 & Poisson's ratio\\
$\rho$ & 2700 kg m$^{-3}$ & Density\\
$\varphi$ & 60$^\circ$ & Orientation of $\sigma_1$\\
$\sigma_n$ & 33.3 MPa & Normal stress on the main fault\\
$R_0$ & 1192 m  & Quasi-static process zone size\\
$ds$ &  79.5 m & Grid size on the main fault\\
$dt$ &  0.11 ms & time step\\
[0.5cm]
Sub-Rayleigh& & \\
\hline
$\tau$ & 13.3 MPa & Shear stress on the main fault \\
$S$ & 1.0 & S ratio\\
$L_c$ & 3092 m  & Nucleation length\\
[0.5cm]
Supershear& & \\
\hline
$\tau$ & 14.5 MPa & Shear stress on the main fault \\
$S$ & 0.7 & S ratio\\
$L_c$ & 2234 m  & Nucleation length\\
\hline

\end{tabular}
\label{tab:planarfaultparameters_material}
\end{table}

\begin{table}
\center
\caption{Variables for contact interactions for modeling a planar fault at 2km depth.}
\vspace{3pt}
On the main fault
\vspace{3pt}

\begin{tabular}{p{2.0cm} p{3cm} p{7.5cm}}
\hline
$f_s$ & 0.6 & Static friction coefficient \\
$f_d$ & 0.2 & Dynamic friction coefficient \\ 
$D_c$ & 0.45 m & Characteristic slip distance\\
$G_{IIC}^{f}$ & 3 MJ m$^{-2}$ & Fracture energy for friction\\
\end{tabular}
\\[0.5cm]
\vspace{3pt}
In the off-fault medium 
\vspace{3pt}

\begin{tabular}{p{2.0cm} p{3cm} p{7.5cm}}
\hline
$f_s$ & 0.6 & Static friction coefficient \\
$f_d$ & 0.2 & Dynamic friction coefficient \\ 
$D_c$ & 1.0 mm & Characteristic slip distance\\
$G_{IC}^{c}$ & 0.7 KJ m$^{-2}$ & Fracture energy for tensile cohesion\\
$G_{IIC}^{c}, G_{IIC}^{f}$ & 5.0 KJ m$^{-2}$ & Fracture energy for shear cohesion and friction\\
$d_{MC}$ & 0.45 & Closeness to failure\\
$C_{I}^p$ & 8.0 MPa & Peak cohesion for opening crack in low cohesion zone\\
$\delta_{I}^{c, c} - \delta_{I}^{c, e}$ & 0.18 mm & Critical displacement for softening of tensile cohesion\\
[0.5cm]
Sub-Rayleigh& & \\
\hline
$C_{II}^p$ & 24.5 MPa & Peak cohesion for shear crack in low cohesion zone\\
$\delta_{II}^{c, c} - \delta_{II}^{c, e}$ & 0.41 mm & Critical displacement for softening of shear cohesion\\
[0.5cm]
Supershear& & \\
\hline
$C_{II}^p$ & 28.1 MPa & Peak cohesion for shear crack in low cohesion zone\\
$\delta_{II}^{c, c} - \delta_{II}^{c, e}$ & 0.35 mm & Critical displacement for softening of shear cohesion\\
\hline
\end{tabular}
\label{tab:planarfaultparameters_contact}
\end{table}

%%%%%%%%%%%%%%%%%%%%%%%%%%%%%%%%%%%%%%%%%%%%%%%%%%%%%%%%%%%%%%%%%%%%%%%%%%%%%%%%

The rupture is artificially nucleated from the nucleation patch, where the peak friction is lower than the initial shear traction on the main fault. The size of nucleation patch $L_c$ is determined by the critical crack length \citep[][]{palmer1973}.
Then it propagates bilaterally on the main fault, dynamically activating off-fault fractures. The x axis is along the fault-parallel direction, while the y axis is along the fault-normal direction. The z axis is thus along depth. Figure \ref{fig:modeldescription}b shows the schematic of case study with depth. We performed a set of 2-D dynamic earthquake rupture modeling to investigate the evolution of coseismic off-fault damage and its effects as a function of depth. We conducted 2-D simulations for depths ranging from z = 2km to 10km in 1km intervals, imposing the corresponding initial stress states as shown in Figure \ref{fig:modeldescription}c. We assume lithostatic condition with depth so that the confining pressure linearly increases with depth. The quasi-static process zone size $R_0$ (see eq. \ref{eq:processzonesizeconstGIIC}) decreases with depth while the fracture energy on the main fault $G_{IIC}^f$ is kept constant (Figure \ref{fig:modeldescription}c). Note that the case study does not address the 3-D effect (e.g. free surface) as we model the dynamic ruptures in plane strain conditions.

For the sake of fair comparison between different depths, the model parameters are nondimensionalized (i.e., made dimensionless) by a combination of scaling factors. $R_0$ [m] and shear wave velocity $c_s$ [m/s] are used to scale the length [m] and the time [s] by $R_0$ and $R_0/c_s$, respectively. Subsequently, other variables are also nondimensionalized by a combination of those two scaling factors. Since the medium's density does not change during the simulations, there is no need for a mass non-dimensionalization in our problem.
%%%%%%%%%%%%%%%%%%%%%%%%%%%%%%%%%%%%%%%%%%%%%%%%%%%%%%%%%%%%%%%%%%%%%%%%%%%%%%%%
\begin{figure}
\center
\noindent\includegraphics[width=\textwidth]{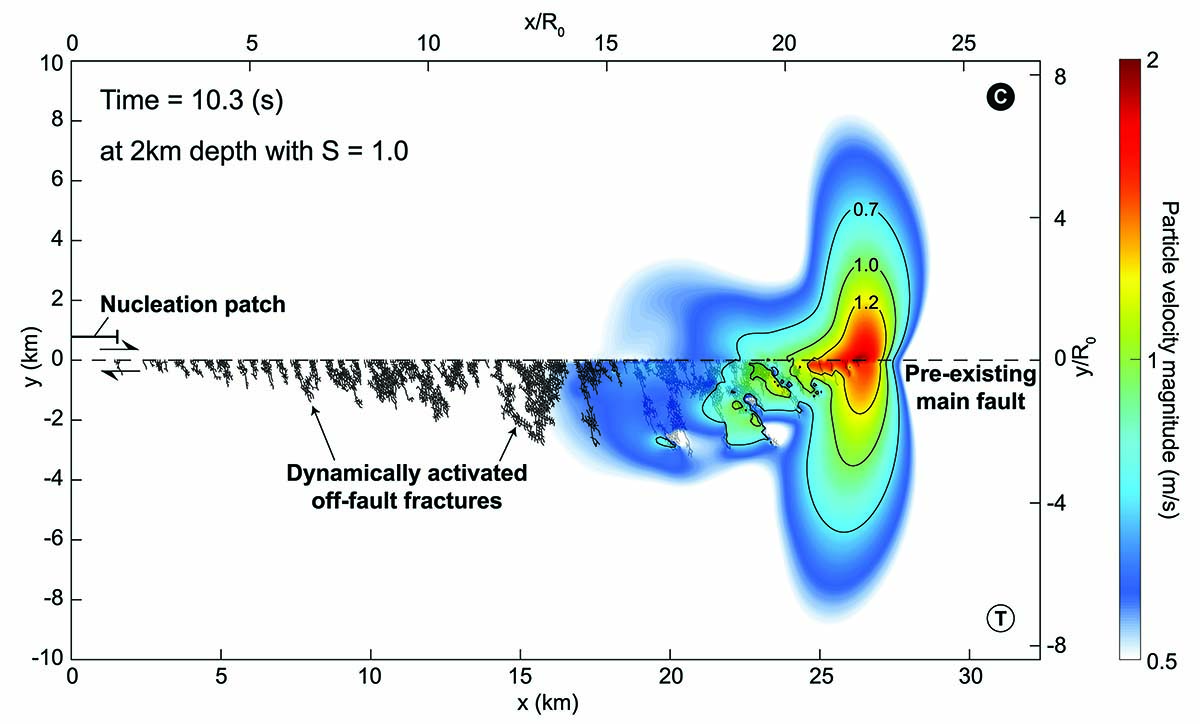}
\caption{Snapshot of the dynamic earthquake rupture with coseismic off-fault damage. We plot only the right part of the model ($x$ $>$ $0$) as the result is symmetrical with respect to the origin. The initial stress state and the material strength correspond to a 2km depth with $S=1.0$. The color contours indicate the particle velocity magnitude. The dotted line indicates the main fault and the solid lines indicate the secondarily activated off-fault fractures. The bottom and left axis show the physical length scales, while the top and right axis show the nondimensionalized lengths scaled by $R_0$. "C" and "T" at right corners indicate compressional and extensional sides of the main fault, respectively.}
\label{fig:snapshotSR}
\end{figure}
%%%%%%%%%%%%%%%%%%%%%%%%%%%%%%%%%%%%%%%%%%%%%%%%%%%%%%%%%%%%%%%%%%%%%%%%%%%%%%%%
\begin{figure}
\center
\noindent\includegraphics[width=\textwidth]{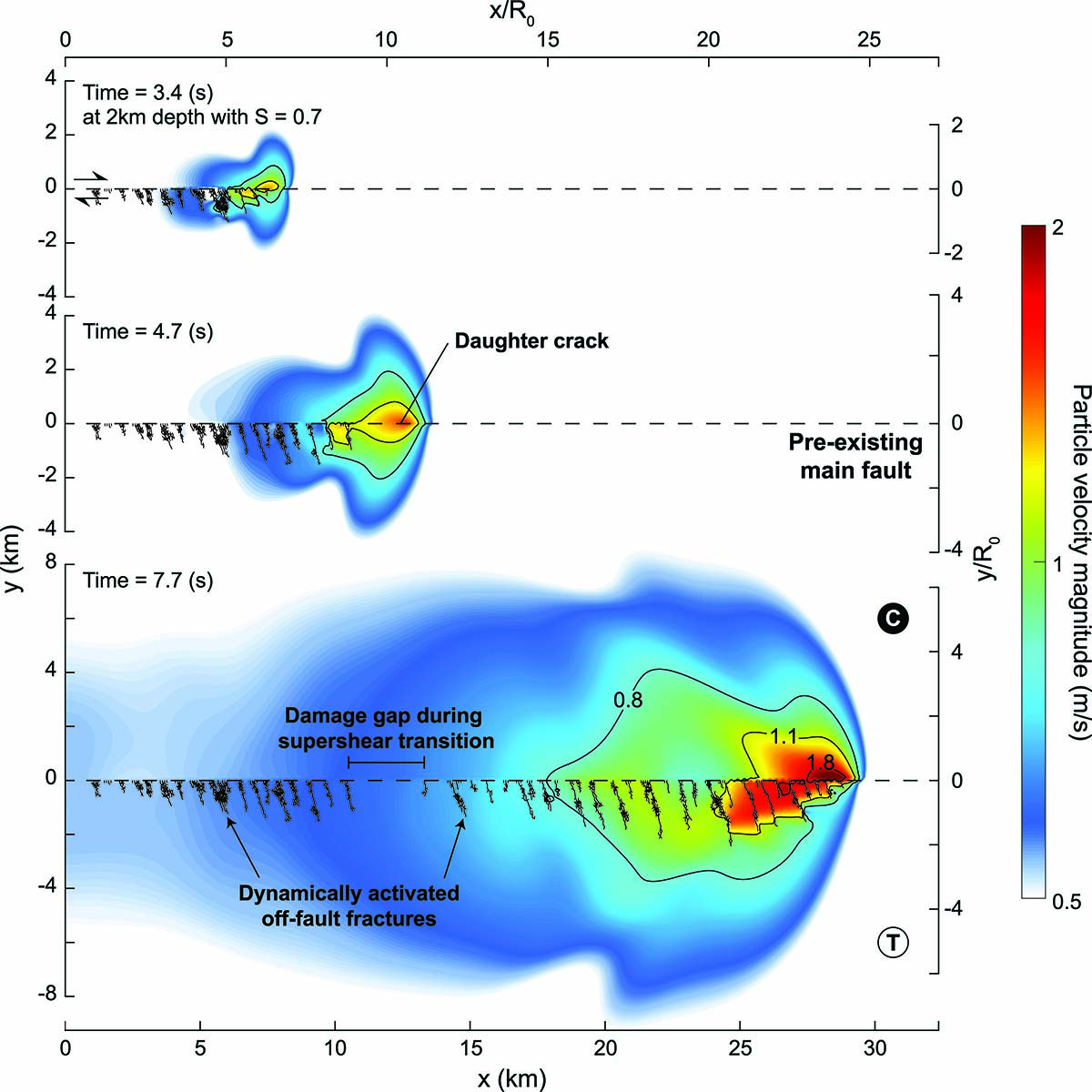}
\caption{Snapshots of supershear rupture at 2km depth with $S=0.7$. The color contours and lines have the same meaning as in Figure \ref{fig:snapshotSR}. The rupture velocity is sub-Rayleigh until $T=3.4$ s (top), then a daughter crack is born ahead of the sub-Rayleigh rupture front at $T=4.7$ s (middle), which transitions to a supershear rupture (bottom). }
\label{fig:snapshotSS}
\end{figure}
%%%%%%%%%%%%%%%%%%%%%%%%%%%%%%%%%%%%%%%%%%%%%%%%%%%%%%%%%%%%%%%%%%%%%%%%%%%%%%%%
Figure \ref{fig:snapshotSR} shows a snapshot of a dynamic earthquake rupture simulation with dynamically activated off-fault fractures, where the particle velocity field and the fracture traces around the main fault are superimposed. The seismic ratio $S$ is equal to 1.0 (see eq. \ref{eq:Sratio}), which results in a sub-Rayleigh rupture. The off-fault fractures are plotted when the traction applied on the potential failure plane (i.e. interfaces between finite elements within the mesh) reaches the cohesive strength and the cohesion starts weakening. Bottom and left axes indicate the fault-parallel and fault-normal distances in physical length scales, while top and right axes indicate the nondimensionalized length scales.

The off-fault fractures are initiated around the rupture tip, forming an intricate fracture network as the main rupture propagates along the main fault. The particle velocity field is significantly perturbed due to the generation of coseismic off-fault damage. The extensional side of the main fault is mostly damaged, which is supported by theoretical analyses of potential failure areas \citep{poliakov2002,rice2005} and other numerical simulation studies \citep[e.g.][]{andrews2005}. The intricate network is formed by means of fracture coalescence between tensile, shear and mixed mode fractures. We later discuss this off-fault fracturing process under a relatively steep angle of the maximum compressive principal stress $\sigma_1$ to the fault ($\psi$ = 60$^{\circ}$), and its effects on the near field radiation and overall energy budget. 

Figure \ref{fig:snapshotSS} shows a set of snapshots for the supershear case with $S = 0.7$. The rupture is nucleated and propagates at sub-Rayleigh speeds in the earlier phase. Then a daughter crack is born ahead of the rupture front at $T = 4.7$ s, which then transitions to a supershear rupture. During the rupture transition from sub-Rayleigh to supershear, characteristic damage patterns appear; there is a gap of coseismic off-fault damage around the transition phase (around x = 12km in Figure \ref{fig:snapshotSS}). This characteristic damage gap has been also pointed out by \citet{templeton2008} and \citet{thomas2018a}. This can be explained by the Lorentz contraction of the dynamic process zone size $R_f(v_r)$. The dynamic process zone size asymptotically shrinks at the rupture's limiting speed, i.e., Rayleigh's wave speed. Hence the damage zone size is minimized around when rupture velocity jumps from sub-Rayleigh to supershear, causing the damage gap in the region. 

%-------------------------------------------------------------------------------
\subsubsection{Rupture velocity}
\begin{figure}
\center
\noindent\includegraphics[width=\textwidth]{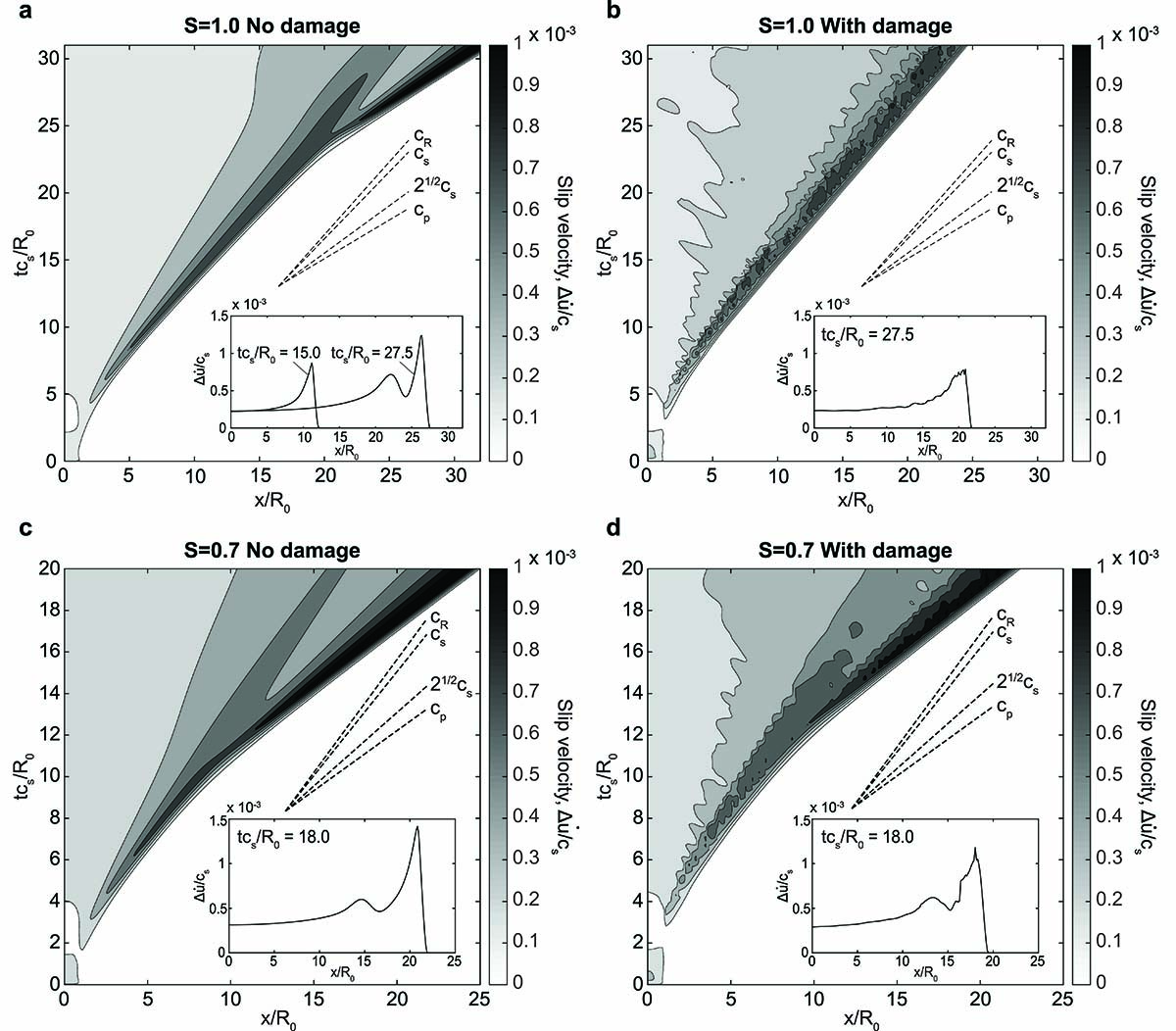}
\caption{The evolution of slip velocity in time and space at 2km depth. There are four cases: (a) $S = 1.0$ with no damage in the off-fault medium (b) $S = 1.0$ with damage (c) $S = 0.7$ with no damage (d) $S = 0.7$ with damage. For the cases without damage, very high cohesion strengths for both tensile and shear failure modes are set so that the off-fault medium behaves as a purely elastic material. 
The grey scale contours indicate the slip velocity. Dotted lines indicate the reference of the slope corresponding to each wave velocity. Insets show the distribution of slip velocity on the main fault at certain time.}
\label{fig:xtplot}
\end{figure}

We next focus on analysing the rupture velocity changes along the main fault. Figure \ref{fig:xtplot} shows the evolution of slip velocity on the main fault for four cases; $S = 1.0$ or 0.7 at 2km depth, each of which with or without considering off-fault damage. For the cases without off-fault damage, the activation of secondary fractures is prevented by imposing very high values of cohesion strength for both tensile and shear modes. Here, we plot the contour of slip velocity in space and time. In Figure \ref{fig:xtplot}a, there is a clear transition from sub-Rayleigh to supershear around $x/R_0 = 20$, which is also shown in the inset. 
However, when the coseismic off-fault damage is taken into account, the supershear transition is not observed during the simulation as shown in Figure \ref{fig:xtplot}b. Hence, the secondary fractures can arrest, or delay, supershear transition in certain stress conditions. This can be explained by the increase in critical slip distance due to the coseismic off-fault damage. The supershear transition length $L^{\text{trans}}$ can be estimated from the Andrews' result \citep{andrews1985,xia2004} as follows
%-------------------------------------------------------------------------------
\begin{equation}
L^{\text{trans}} = \dfrac{1}{9.8(S_{\text{crit}} - S)^3} \dfrac{1+\nu}{\pi}\dfrac{\tau^p - \tau^r}{(\tau - \tau^r)^2} \mu D_c,
\label{eq:sstransition}
\end{equation}
%-------------------------------------------------------------------------------
where $S_{\text{crit}}$ is the threshold for the supershear transition ($S_{\text{crit}} = 1.77$ for 2-D), $\nu$ is the Poisson's ratio of the medium, $\tau_p$, $\tau_r$ and $\tau$ are peak friction (eq. \ref{eq:peakfric}), residual friction (eq. \ref{eq:residualfric}) and shear traction on the fault, respectively, $\mu$ is the shear modulus and $D_c$ is the critical slip distance for friction (eq. \ref{eq:Dcwithdepth}). 
$D_c$ is initially uniform on the main fault. However, the effective critical slip distance, which takes into account the fracture energy associated with both on and off the fault, increases with the evolution of coseismic off-fault damage (see \citet{okubo2018h}). Therefore, $L^{\text{trans}}$ also increases as it is proportional to Dc.

%-------------------------------------------------------------------------------
%-------------------------------------------------------------------------------
\begin{figure}
\center
\noindent\includegraphics[width=0.8\textwidth]{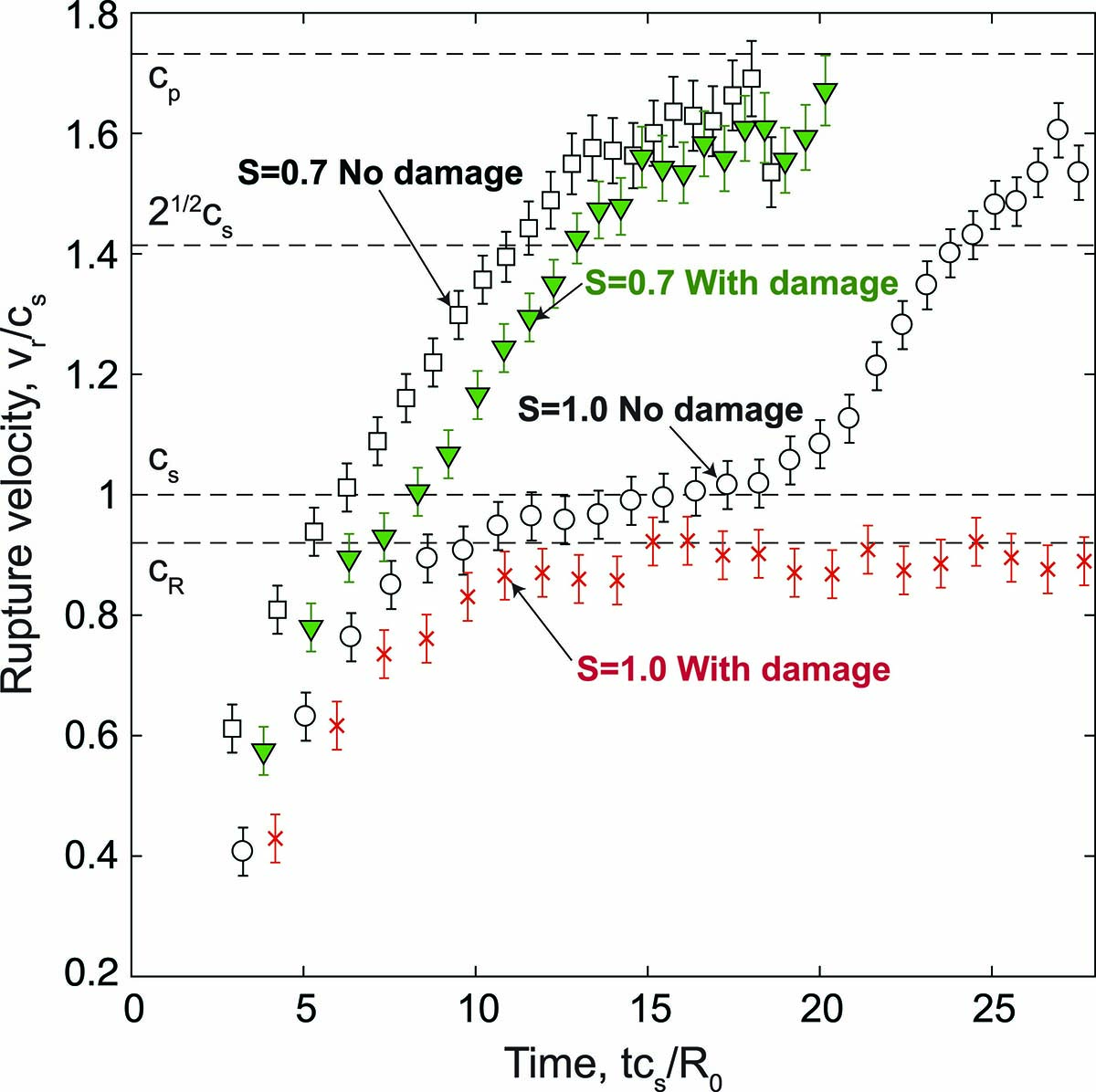}
\caption{Rupture velocity inferred from Figure \ref{fig:xtplot}. Due to inherent discretization errors, it is difficult to precisely capture the jump of rupture velocity from sub-Rayleigh to supershear. The error is estimated from the difference between the slope of $c_R$ and $c_s$, the grid spacing and the sampling rate of slip velocity.}
\label{fig:rupturevelocity}
\end{figure} 
%-------------------------------------------------------------------------------
Figures \ref{fig:xtplot}c and \ref{fig:xtplot}d show the cases with $S=0.7$, where the rupture transitions to supershear for both cases with and without off-fault damage because of the large contrast of the initial shear traction to the normal traction on the main fault. The time at which the supershear transition happens is delayed for the cases with off-fault damage due to the decrease of rupture velocity, whereas the difference of transition length is still obscure with these results. The two insets shown in figures \ref{fig:xtplot}c and \ref{fig:xtplot}d show a clear difference in the slip velocity peak and the fluctuations. In addition, the rupture arrival is delayed by the coseismic off-fault damage, implying a decrease in rupture velocity.

The rupture velocity is calculated based on first arrival times along the main fault. Figure \ref{fig:rupturevelocity} shows the evolution of rupture velocity as a function of time. We take the time derivatives of first arrival time in discretized space along the main fault to calculate the representative rupture velocity at a certain position. Since it is difficult to capture the exact time when the rupture velocity jumps to supershear, which is where the curve of first arrival time has a kink and is non-differentiable, the error caused by the smoothing of the rupture velocity is taken into account as shown by the error bars in Figure \ref{fig:rupturevelocity}. Therefore, the markers in the forbidden zone $c_R < v_R <c_s$ do not conclusively indicate that the rupture velocity is between them due to the uncertainty in the method used to calculate it. 

Regardless of the uncertainty, the comparison between the cases with and without off-fault damage shows the effects of coseismic off-fault damage on the rupture velocity and on the supershear transition. The rupture transitions to supershear for both cases with $S=0.7$, though the rate of increase in rupture velocity is lower for the case with off-fault damage. However, the supershear transition is suppressed due to the coseismic off-fault damage with $S=1.0$.

Further discussion on the rupture dynamics with coseismic off-fault damage using an infinite planar fault model can be found in \citet{okubo2018h} and \citet{okubo2019}. In this subsection, we summarized the modeling with this fault model as it is fundamental for the rupture modeling with first-order geometrical complexity demonstrated in the following section.

%%%%%%%%%%%%%%%%%%%%%%%%%%%%%%%%%%%%%%%%%%%%%%%%%%%%%%%%%%%%%%%%%%%%%%%%%%%%%%%%
\subsection{Rupture along a kink}
%%%%%%%%%%%%%%%%%%%%%%%%%%%%%%%%%%%%%%%%%%%%%%%%%%%%%%%%%%%%%%%%%%%%%%%%%%%%%%%%
Next, we model dynamic earthquake rupture along a fault kink. Since the stress is locally concentrated due to the fault kink, the coseismic off-fault damage could be enhanced around the kink. We thus conducted dynamic earthquake rupture modeling with the fault kink, which bends towards either the compressional or extensional side of the fault.
%-------------------------------------------------------------------------%
\begin{figure}[ht!]
\centering
\noindent\includegraphics[width=0.7\textwidth]{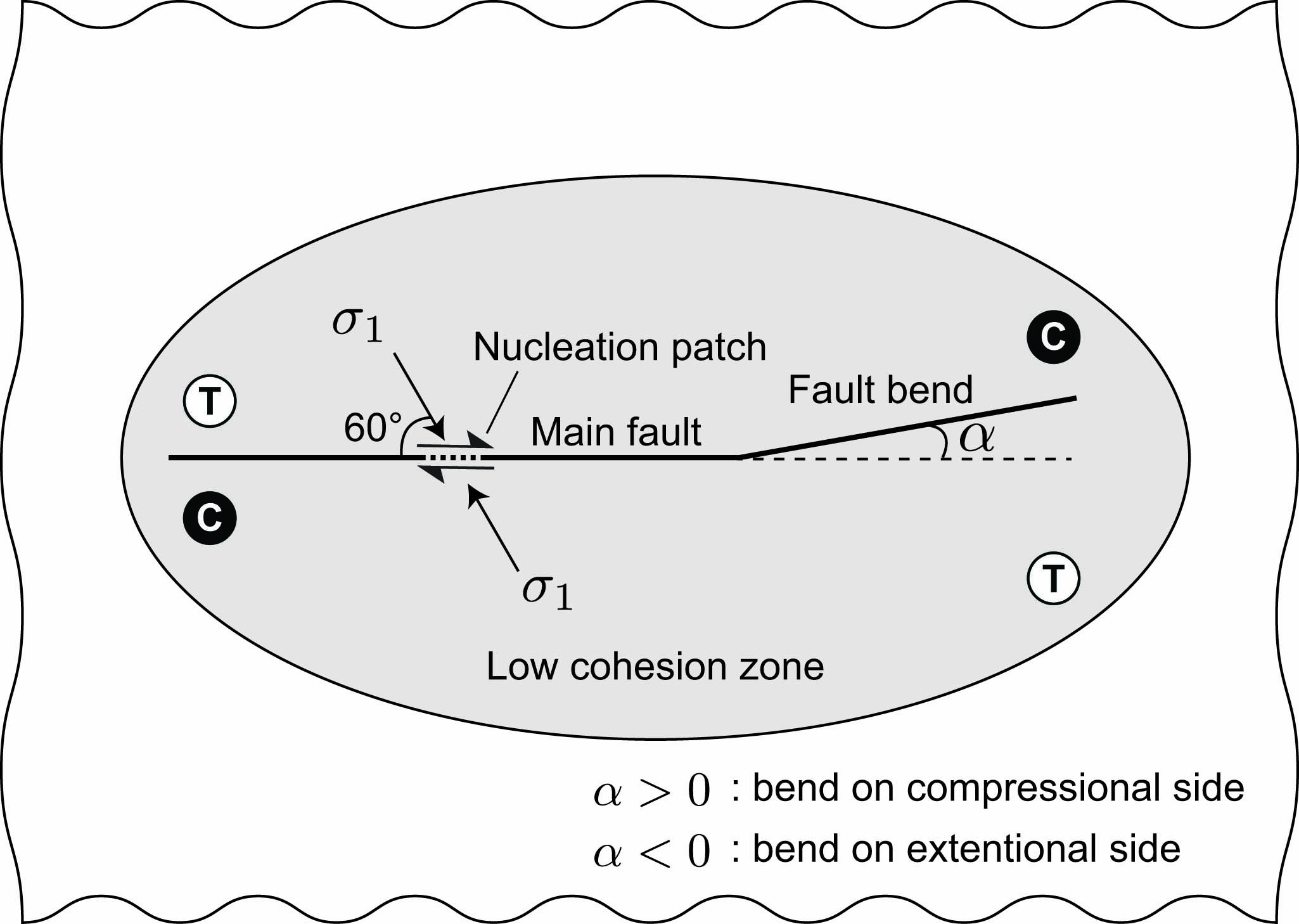}
\caption{Model setup for a fault kink. The angle of the fault bend is $\alpha$. The material constants and relevant model parameters are listed in Table \ref{tab:faultkinkparameters_matconst} and \ref{tab:faultkinkparameters_contact}. C and T indicate the compressive and extensional sides, respectively. The length of the main fault is 12.5km, while the fault bend is 7.5km.}
\label{fig:kinkfaultschematic}
\end{figure}

%-------------------------------------------------------------------------%
\begin{table}
\center
\caption{Material constants and parameters for modeling a fault kink, step-over and roughness.}
\begin{tabular}{p{2.0cm} p{3cm} p{7.5cm}}
\hline
\textbf{Valuables} & \textbf{Values} & \textbf{Descriptions} \\ \hline
$E$ & 75 GPa & Young's modulus\\
$\mu$ & 30 GPa & Shear modulus\\
$\nu$ & 0.25 & Poisson's ratio\\
$\rho$ & 2700 kg m$^{-3}$ & Density\\

$\sigma_n$ & 40 MPa & Normal stress on the main fault\\
$\tau$ & 16 MPa & Shear stress on the main fault \\
$\sigma_1$ & 49 MPa & Maximum compressive principal stress\\
$\sigma_2$ & 12 MPa & Minimum compressive principal stress\\
$S$ & 1.0 & S ratio\\
$d_{MC}$ & 0.45 & Closeness to failure\\
$\varphi$ & 60 $^\circ$ & Orientation of $\sigma_1$\\
$R_0$ & 552 m  & Quasi-static process zone size\\
$L_c$ & 1200 m  & Nucleation length\\
$ds$ & 55 m & Grid size on the main fault\\
\hline
\end{tabular}

\label{tab:faultkinkparameters_matconst}
\end{table}

\begin{table}
\center
\caption{Variables for contact interactions for a fault kink, step-over and roughness.}
\vspace{3pt}
On the main fault
\vspace{3pt}

\begin{tabular}{p{2.0cm} p{3cm} p{7.5cm}}
\hline
$f_s$ & 0.6 & Static friction coefficient \\
$f_d$ & 0.2 & Dynamic friction coefficient \\ 
$D_c$ & 0.25 m & Characteristic slip distance\\
$G_{IIC}^{f}$ & 2 MJ m$^{-2}$ & Fracture energy for friction\\
\end{tabular}
\\[0.5cm]
\vspace{3pt}
In the off-fault medium 
\vspace{3pt}

\begin{tabular}{p{2.0cm} p{3cm} p{7.5cm}}
\hline
$f_s$ & 0.6 & Static friction coefficient \\
$f_d$ & 0.2 & Dynamic friction coefficient \\ 
$D_c$ & 12.5 mm & Characteristic slip distance\\
$G_{IC}^{c}$ & 8 KJ m$^{-2}$ & Fracture energy for tensile cohesion\\
$G_{IIC}^{c}, G_{IIC}^{f}$ & 90 KJ m$^{-2}$ & Fracture energy for shear cohesion and friction\\
$C_{I}^p$ & 8 MPa & Peak cohesion for opening crack in low cohesion zone\\
$C_{II}^p$ & 30 MPa & Peak cohesion for shear crack in low cohesion zone\\
$\delta_{I}^{c, c} - \delta_{I}^{c, e}$ & 2.0 mm & Critical displacement for softening of tensile cohesion\\
$\delta_{II}^{c, c} - \delta_{II}^{c, e}$ & 6.0 mm & Critical displacement for softening of shear cohesion\\
\hline
\end{tabular}
\label{tab:faultkinkparameters_contact}
\end{table}

Figure \ref{fig:kinkfaultschematic} shows the model setup for a fault kink. The model parameters and are listed in Tables \ref{tab:faultkinkparameters_matconst} and \ref{tab:faultkinkparameters_contact}. The angle of bend, $\alpha$, is an important parameter for the rupture propagation on the fault kink. When $\alpha > 0$, the fault bends on the compressional side of the main fault. In this case, the ratio of shear traction to the normal traction, $\tau/\sigma_n$, decreases as $\alpha$ increases. Thus the rupture is less likely to propagate along a fault bend with large $\alpha$. On the other hand, when $\alpha < 0$, the fault bends on the extensional side of the main fault, where the $\tau/\sigma_n$ is larger on the fault bend than on the main fault.

\begin{figure}[ht!]
\centering
\noindent\includegraphics[width=\textwidth]{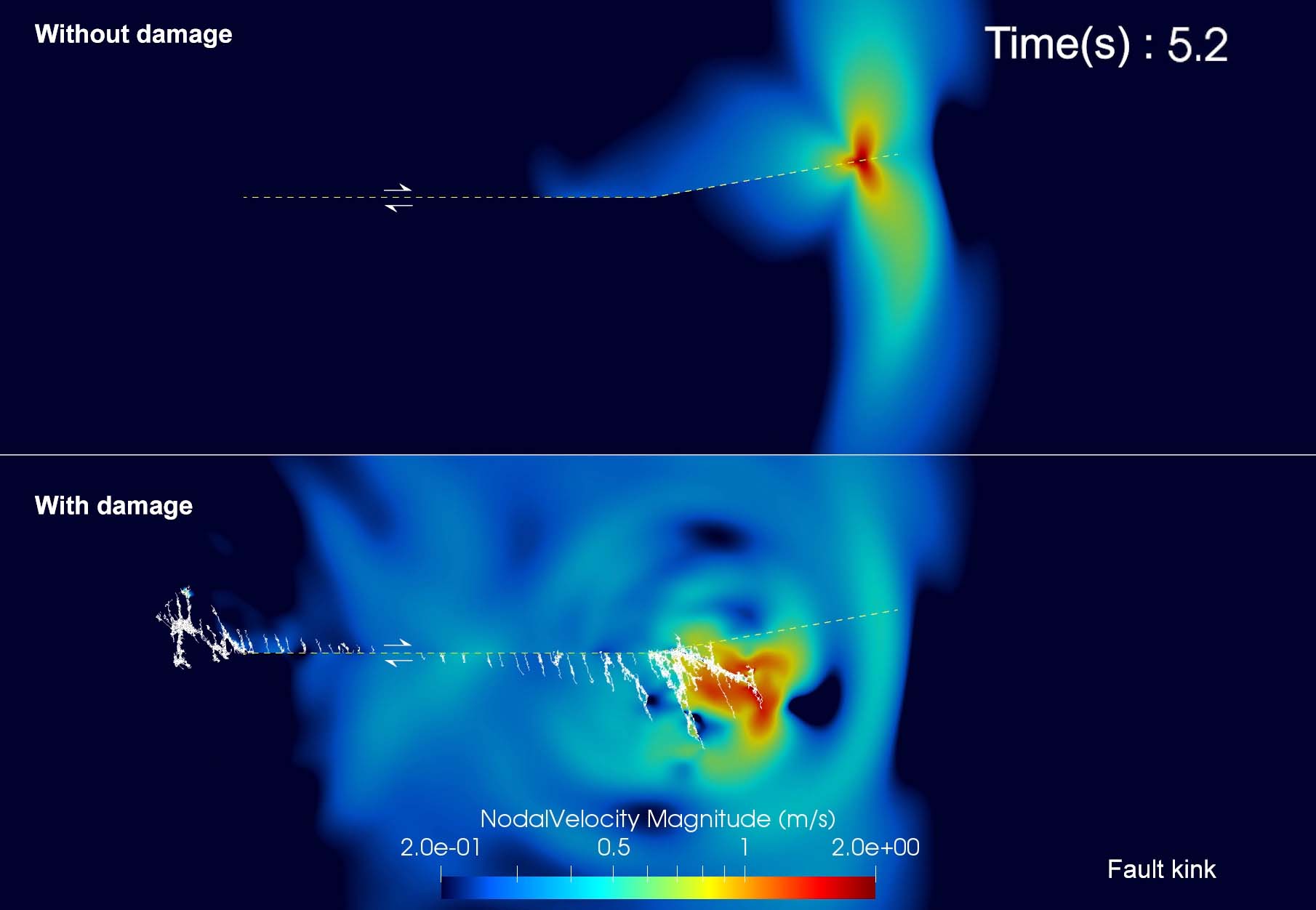}
\caption[Snapshot for fault kink bent on the compressional side of the main fault at $T = 5.2$s.]{Snapshot at $T = 5.2$s. The rupture propagates on the prescribed bent fault for the case without off-fault damage, whereas the rupture does not propagate the prescribed fault with the coseismic off-fault damage. Instead of the arrest of the rupture at kink, a new fault branch is generated toward the extensional side.}
\label{fig:kinkfault_c_snap03}
\end{figure}

\begin{figure}[ht!]
\centering
\noindent\includegraphics[width=\textwidth]{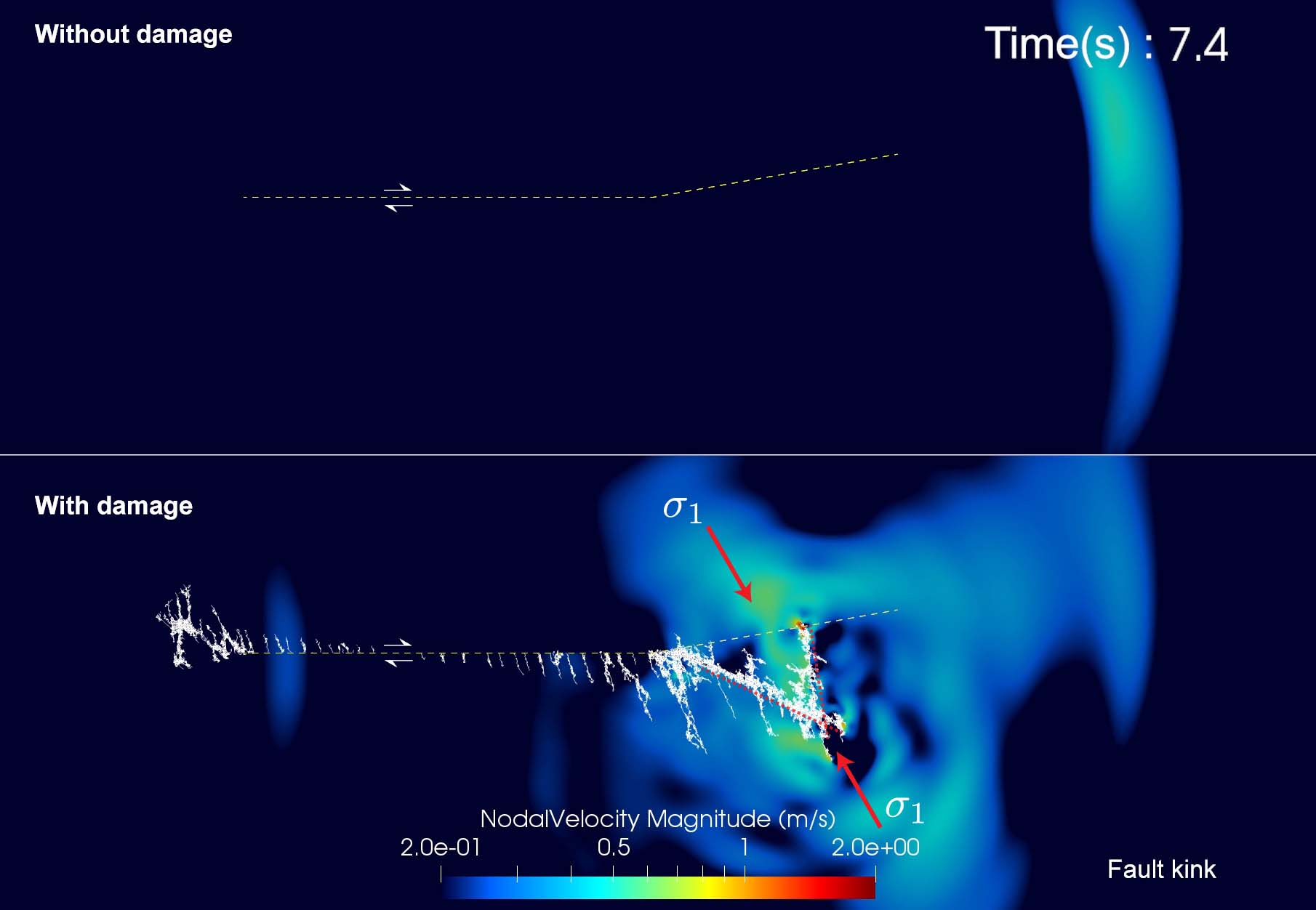}
\caption{Snapshot at T = 7.4s. Eventually the rupture is arrested at the edge of the bent fault for the case without off-fault damage. For the case with allowing for the coseismic off-fault damage, however, the branch grows as activating a lot of off-fault damage from the fault kink, and induces the secondary branch upward as guided by red dashed lines. }
\label{fig:kinkfault_c_snap04}
\end{figure}

Here, we demonstrate the cases for $\alpha = +10^\circ$ with and without coseismic off-fault damage to investigate the rupture dynamics and the associated damage patterns (see \citet{okubo2018h} for the case with $\alpha = -10^\circ$).  Figures \ref{fig:kinkfault_c_snap03} and \ref{fig:kinkfault_c_snap04} show the results obtained for a fault kink bent on the compressional side of the main fault. In this case, the rupture is less likely to propagate along the fault bend due to the decrease in $\tau/\sigma_n$. Nevertheless, in the case without off-fault damage, the rupture propagates completely along the prescribed fault. This result is in accordance with \citet{templeton2009}. However, in the case with coseismic off-fault damage, the rupture is arrested at the kink, while a significant amount of damage is caused on the extensional side, resulting in the formation of a secondary fault branch. Eventually, two major fracture paths are generated; the orientation of these branches corresponding to the conjugate failure planes of $\sigma_1$ (Figure \ref{fig:kinkfault_c_snap04}). Therefore, we expect that secondary fault branches from kinks are naturally generated corresponding to the conjugate failure planes of the regional stress.

%-------------------------------------------------------------------------%

%-------------------------------------------------------------------------%

%%%%%%%%%%%%%%%%%%%%%%%%%%%%%%%%%%%%%%%%%%%%%%%%%%%%%%%%%%%%%%%%%%%%%%%%%%%%%%%%
\subsection{Rupture on a step-over fault system}
\begin{figure}
\centering
\noindent\includegraphics[width=0.8\textwidth]{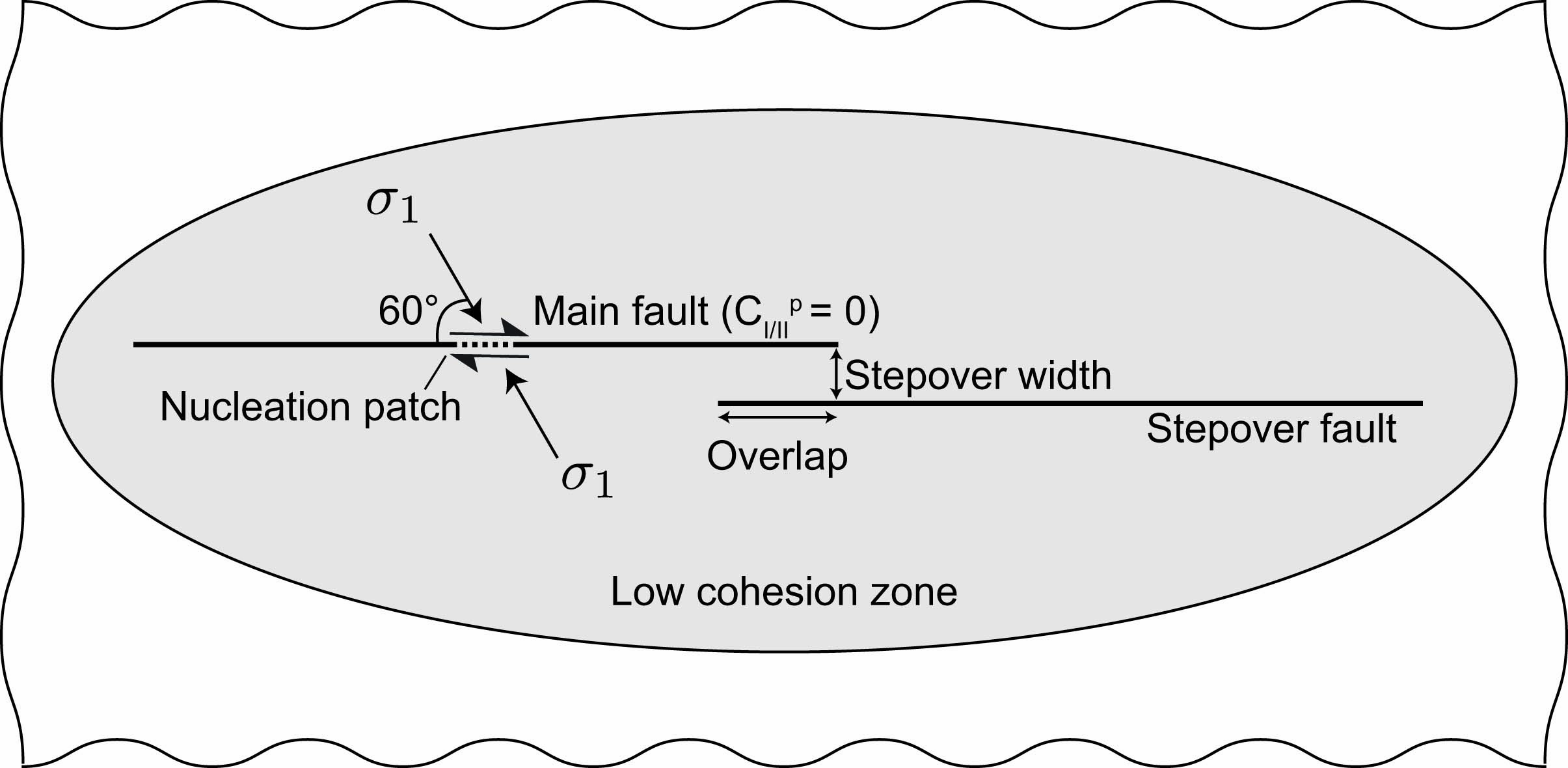}
\caption[Model setup for modeling  faults.]{Model setup for modeling  faults. The material constants and relevant model parameters are same with the fault kink model (Table \ref{tab:faultkinkparameters_matconst} and \ref{tab:faultkinkparameters_contact}). The fault length is 15km, the  width is 600m ($\sim$ 0.5$L_c$) and the overlap is 2.5km ($\sim$ 2.1$L_c$).}
\label{fig:stepovermodelsetup}
\end{figure}

%%%%%%%%%%%%%%%%%%%%%%%%%%%%%%%%%%%%%%%%%%%%%%%%%%%%%%%%%%%%%%%%%%%%%%%%%%%%%%%%
The step-over faults are another important component of natural fault networks. What is of interest here is determining whether the rupture will jump from the main fault to the step-over fault near the main fault. Systematic numerical experiments of step-over faults, pioneered by \citet{harris1991}, demonstrated the geometrical conditions to nucleate the secondary rupture on the step-over fault. Parallel strike-slip faults are widely used as an example of simple step-over faults (Figure \ref{fig:stepovermodelsetup}). The relative position of the  fault with respect to the main fault is controlled by two parameters:  width and overlap. The ability of the main rupture jumping onto the  fault segments depends on whether the  fault is located on the compressional side (compressional step) or the extensional side (dilational step) of the main fault. \citet{harris1991} shows that the dilational step is more likely to induce a second rupture on the  fault. Thus in this section, we demonstrate the dilation step, and compare the cases with and without off-fault damage to investigate the effects of coseismic off-fault damage on the rupture dynamics for the  faults.
%-------------------------------------------------------------------------%

%-------------------------------------------------------------------------%
\begin{figure}[tp]
\centering
\noindent\includegraphics[width=\textwidth]{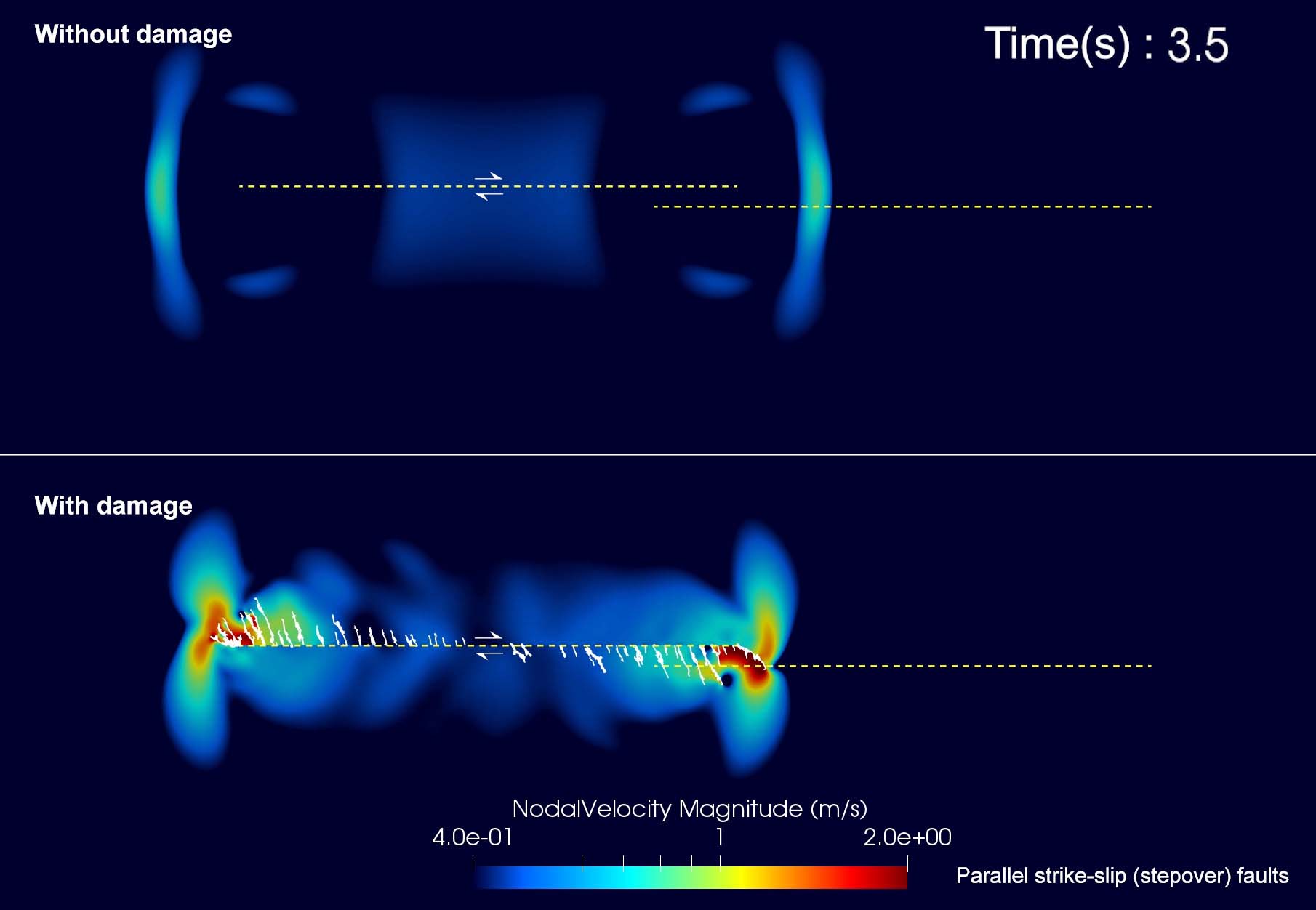}
\caption{Snapshot at T = 3.5s. The rupture reaches the edges of the main fault, and is arrested for the case without off-fault damage, whereas the coseismic off-fault damage grows toward the step-over fault.}
\label{fig:stepover_snap02}
\end{figure}

%-------------------------------------------------------------------------%

\begin{figure}[tp]
\centering
\noindent\includegraphics[width=\textwidth]{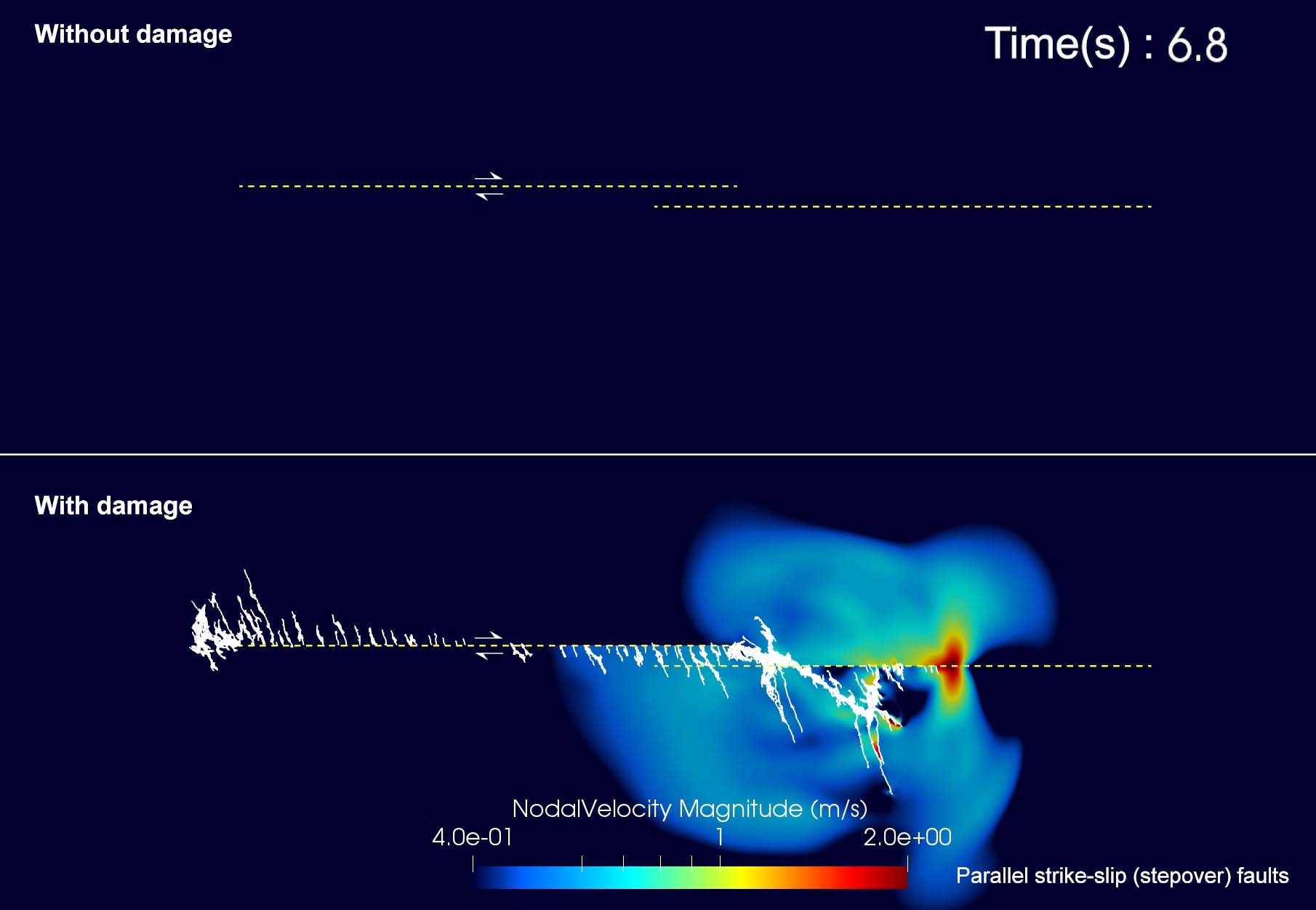}
\caption{Snapshot at T = 6.8s. The secondary rupture is eventually nucleated on the step-over fault due to the off-fault damage.}
\label{fig:stepover_snap03}
\end{figure}

%-------------------------------------------------------------------------%

Figures \ref{fig:stepover_snap02} and \ref{fig:stepover_snap03} show two snapshots of the results obtained for a dilational step case both with and without coseismic off-fault damage. The rupture is nucleated in the middle of the main fault and propagates bilaterally. The  is set as 600m ($\sim$ 0.5$L_c$), and the overlap is 2.5km ($\sim$ 2.1$L_c$). When the rupture reaches the edges of the main fault, the rupture is arrested for the case without coseismic off-fault damage, and does not induce a second rupture on the  fault (Figure \ref{fig:stepover_snap02}). However, when off-fault damage is considered major off-fault fracture paths evolve from the right edge of the main fault, which reach the  fault. Then as the coseismic off-fault damage evolves around the edge of the main fault, the secondary rupture is nucleated close to the major damage zone on the  fault (Figure \ref{fig:stepover_snap03}). Since the secondary rupture is not nucleated for the case without coseismic off-fault damage, implying this combination of fault geometry and initial stress conditions is not favorable for "rupture jumping", we conclude that the coseismic off-fault damage around the  fault can increase the probability of rupture jump onto the  fault. These preliminary results for the  faults demonstrate the need of parametric studies to rectify the conditions of rupture jumps with coseismic off-fault damage.
%%%%%%%%%%%%%%%%%%%%%%%%%%%%%%%%%%%%%%%%%%%%%%%%%%%%%%%%%%%%%%%%%%%%%%%%%%%%%%%%

\subsection{Rupture along a rough fault}
%%%%%%%%%%%%%%%%%%%%%%%%%%%%%%%%%%%%%%%%%%%%%%%%%%%%%%%%%%%%%%%%%%%%%%%%%%%%%%%%
In the previous sections, we only consider a combination of planar faults even though it has a kink, or step over faults. However, it is recognized that fault roughness also plays a crucial role in rupture dynamics, radiations and coseismic off-fault damage \citep[e.g.][]{dunham2011b}. In this section, we demonstrate a preliminary result with a self-similar fault to investigate the rupture processes on the rough faults with coseismic off-fault damage. The self-similar fault geometry is reproduced based on \citet{dunham2011b}.  The self-similar fault profile has a spectral density, $P_m (k)$, as follows
\begin{equation}
P_{m} (k) = (2\pi)^3 \beta ^2 |k|^{-1},
\end{equation}
where $k$ is the wave number and $\beta$ is a parameter to determine the magnitude of fluctuation of the fault.
%-------------------------------------------------------------------------%
\begin{figure}[ht!]
\centering
\noindent\includegraphics[width=\textwidth]{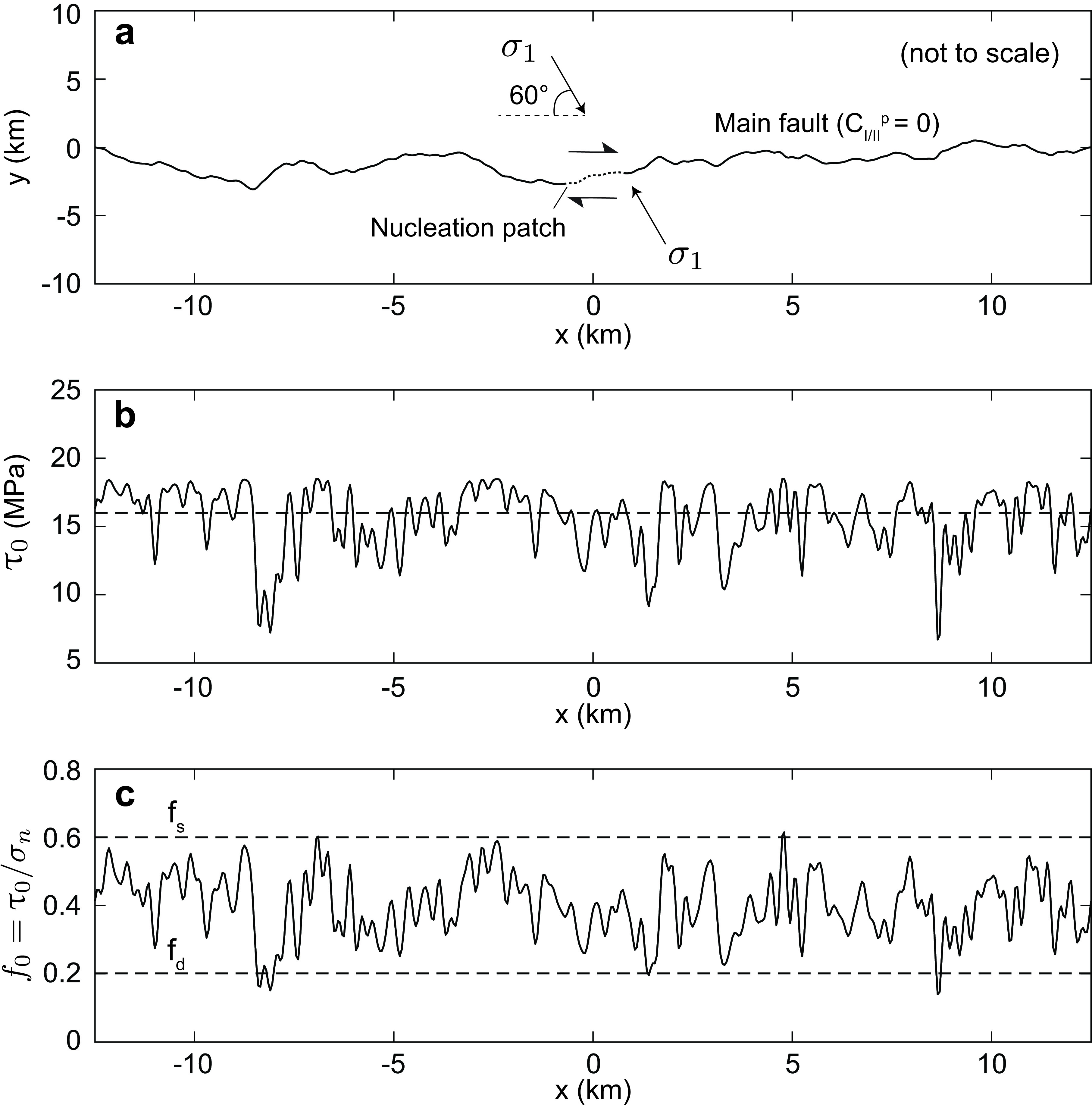}
\caption{Self-similar fault geometry, initial shear traction and initial $f_0$ on the main fault. (a) Fault geometry. The nucleation patch is set in the middle of the fault, which is much larger than $L_c$ to nucleate the rupture on the rough fault. Model parameters are same with the fault kink model (Table \ref{tab:faultkinkparameters_matconst} and \ref{tab:faultkinkparameters_contact}). The fault length is finite, set as 25km, and the entire fault is encompassed by the low cohesion zone. (b) Initial shear traction on the main fault. The dashed line indicates the reference shear traction with a planar fault (16MPa). (c) Initial ratio of $\tau_0$ to $\sigma_n$. The dashed lines indicate the static and dynamic friction coefficients on the main fault.}
\label{fig:roughfaultinfo}
\end{figure}
%-------------------------------------------------------------------------%
Figure \ref{fig:roughfaultinfo} shows the self-similar fault geometry and the shear traction on the fault. $\beta$ is to 3.2 X 10$^{-3}$. We chose the fault geometry so that the initial ratio of $\tau_0$ to $\sigma_n$, $f_0$, is globally less than the static friction coefficient, $f_s$ to avoid unexpected rupture nucleation during the loading phase.
%-------------------------------------------------------------------------%
\begin{figure}[ht!]
\centering
\noindent\includegraphics[clip,width=\textwidth]{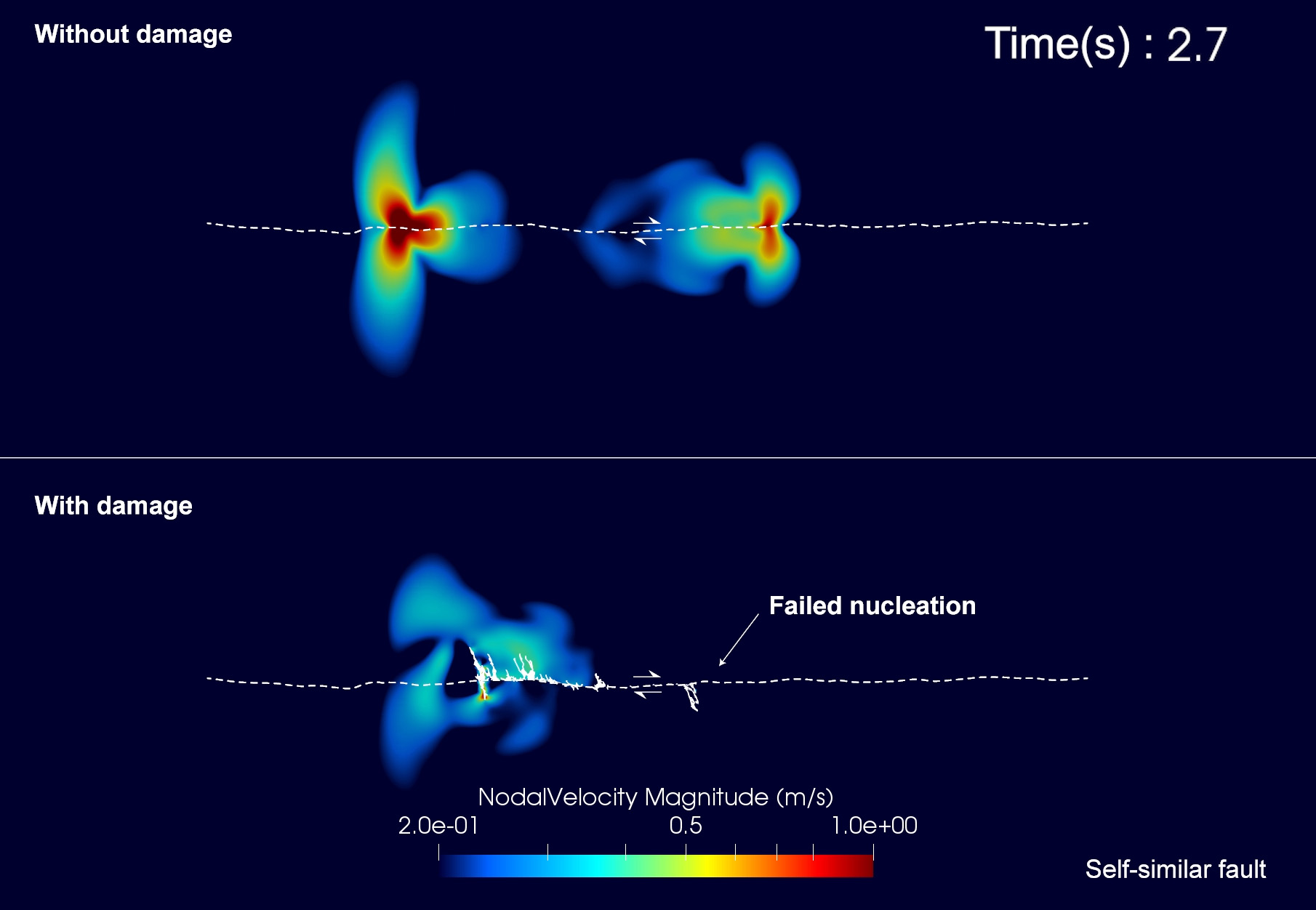}
\caption[Snapshot for self-similar fault at $T = 2.7$s.]{The dashed line indicates the prescribed main fault. Top window show the result without coseismic off-fault damage, while the bottom window show the result with off-fault damage. The arrows indicate the sense of slip on the main fault. The white lines indicate the secondarily activated off-fault fractures.}
\label{fig:roughfault_snap01}
\end{figure}
%-------------------------------------------------------------------------%
\begin{figure}[ht!]
\centering
\noindent\includegraphics[clip,width=\textwidth]{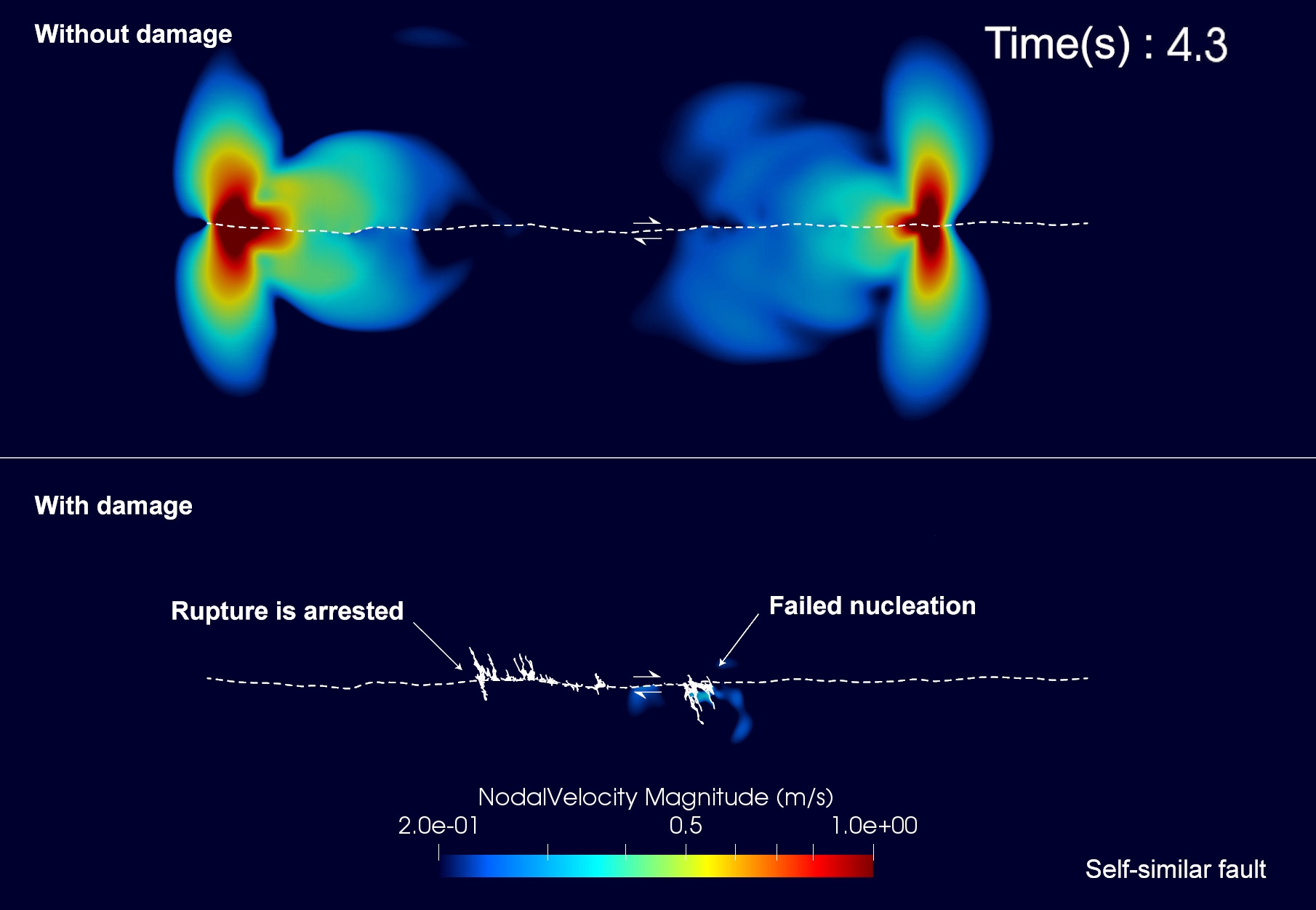}
\caption[Snapshot for self-similar fault at $T = 4.3$s.]{Snapshot at $T = 4.3$s. The rupture on the left side for the case with off-fault damage is also arrested due to the off-fault damage. }
\label{fig:roughfault_snap03}
\end{figure}
%-------------------------------------------------------------------------%

Figures \ref{fig:roughfault_snap01} and \ref{fig:roughfault_snap03} show snapshots for the cases with and without off-fault damage. In the case without off-fault damage, the rupture is nucleated and propagates bilaterally. However, in the case with off-fault damage, the rupture is not successfully nucleated on the right side of the nucleation patch (Figure \ref{fig:roughfault_snap01}) due to prominent cracking at the edges of nucleation patch. 
One of the reasons for the nucleation failure is the artificial manipulation of nucleation process, where low $f_s$ is assigned within the nucleation patch, which causes abrupt change of $f_s$ at the edges of nucleation patch and consequent stress concentrations. Thus the nucleation process needs to be reconsidered to nucleate rupture on the rough fault. 

In addition, the rupture on the left side is also arrested by the coseismic off-fault cracks (Figure \ref{fig:roughfault_snap03}). Thus the main fault is not fully ruptured for the case with coseismic off-fault damage. Therefore, we conclude that the roughness tends to arrest the earthquake ruptures due to the coseismic off-fault damage.
Further parametric studies are needed to investigate the condition of the arrest of ruptures and the associated off-fault damage patterns on the rough fault.

%%%%%%%%%%%%%%%%%%%%%%%%%%%%%%%%%%%%%%%%%%%%%%%%%%%%%%%%%%%%%%%%%%%%%%%%%%%%%%%%
\section{Conclusion}
We demonstrated the continuum-discontiuum approach using FDEM, showing the capability of modeling dynamic earthquake rupture with the coseismic activation of localized off-fault fractures and its effect on the rupture dynamics. Although the number of parameters to define the constitutive law and the contact interactions increases in comparison to the canonical FEM, we can fairly constrain those parameters following a set of formulations presented in the section \ref{sec:methodorogyintro}. The result shows the non-negligible effect of coseismic off-fault damage on the rupture dynamics such as the decrease in rupture velocity.

Each first-order complex fault geometry has a unique coseismic off-fault damage pattern, which would help elucidate the complicated earthquake rupture scenario on the real fault system. We need to further explore these fundamental fault system to address the following questions:
\begin{enumerate}
	\item What is the critical angle associated with fault kink, which decides whether the rupture propagates on the pre-existing fault or generates a dominant fracture path in the orientation of conjugate failure plane?
	\item How does the off-fault damage modify the criteria of rupture jump onto the step-over fault?  
	\item How is the off-fault damage pattern and residual stress on the rough fault?	 
\end{enumerate}
This paper demonstrates that FDEM is an appropriate numerical approach to conduct parametric studies on fault systems, such as those described above, in order to answer key standing questions in the earthquake rupture arena.


\begin{thebibliography}{47}
\providecommand{\natexlab}[1]{#1}
\providecommand{\url}[1]{{#1}}
\providecommand{\urlprefix}{URL }
\expandafter\ifx\csname urlstyle\endcsname\relax
  \providecommand{\doi}[1]{DOI~\discretionary{}{}{}#1}\else
  \providecommand{\doi}{DOI~\discretionary{}{}{}\begingroup
  \urlstyle{rm}\Url}\fi
\providecommand{\eprint}[2][]{\url{#2}}

\bibitem[{Andrews(1976)}]{andrews1976}
Andrews DJ (1976) Rupture velocity of plane strain shear cracks. J Geophys Res
  81(B32):5679--5689, \doi{10.1029/JB081i032p05679}

\bibitem[{Andrews(1985)}]{andrews1985}
Andrews DJ (1985) Dynamic plane-strain shear rupture with a slip-weakening
  friction law calculated by a boundary integral method. Bull Seism Soc Am
  75:1--21

\bibitem[{Andrews(2005)}]{andrews2005}
Andrews DJ (2005) Rupture dynamics with energy loss outside the slip zone. J
  Geophys Res 110:1307, \doi{10.1029/2004JB003191}

\bibitem[{Aochi and Fukuyama(2002)}]{aochi2002a}
Aochi H, Fukuyama E (2002) Three-dimensional nonplanar simulation of the 1992
  landers earthquake. J Geophys Res 107(B2), \doi{10.1029/2000JB000061}

\bibitem[{Chester et~al.(1993)Chester, Evans, and Biegel}]{chester1993}
Chester FM, Evans JP, Biegel RL (1993) Internal structure and weakening
  mechanisms of the san andreas fault. J Geophys Res 98:771--786,
  \doi{10.1029/92JB01866}

\bibitem[{Cho et~al.(2003)Cho, Ogata, and Kaneko}]{cho2003}
Cho SH, Ogata Y, Kaneko K (2003) {Strain-rate dependency of the dynamic tensile
  strength of rock}. Int J Rock Mech Min Sci 40(5):763--777,
  \doi{10.1016/S1365-1609(03)00072-8}

\bibitem[{De~La~Puente et~al.(2009)De~La~Puente, Ampuero, and
  K{\"a}ser}]{delapuente2009}
De~La~Puente J, Ampuero JP, K{\"a}ser M (2009) Dynamic rupture modeling on
  unstructured meshes using a discontinuous {Galerkin method}. J Geophys Res
  114(B10):B10302, \doi{10.1029/2008JB006271}

\bibitem[{Dunham et~al.(2011)Dunham, Belanger, Cong, and Kozdon}]{dunham2011b}
Dunham EM, Belanger D, Cong L, Kozdon JE (2011) Earthquake ruptures with
  strongly rate-weakening friction and off-fault plasticity, part 2: Nonplanar
  faults. Bull Seism Soc Am 101(5):2308--2322, \doi{10.1785/0120100076}

\bibitem[{Euser et~al.(2019)Euser, Rougier, Lei, Knight, Frash, Carey,
  Viswanathan, and Munjiza}]{euser2019}
Euser B, Rougier E, Lei Z, Knight EE, Frash LP, Carey JW, Viswanathan H,
  Munjiza A (2019) Simulation of fracture coalescence in granite via the
  combined finite--discrete element method. Rock Mechanics and Rock Engineering
  52(9):3213--3227, \doi{10.1007/s00603-019-01773-0}

\bibitem[{Faulkner et~al.(2011)Faulkner, Mitchell, Jensen, and
  Cembrano}]{faulkner2011}
Faulkner DR, Mitchell TM, Jensen E, Cembrano J (2011) Scaling of fault damage
  zones with displacement and the implications for fault growth processes. J
  Geophys Res 116(B05403), \doi{10.1029/2010JB007788}

\bibitem[{Fletcher et~al.(2014)Fletcher, Teran, Rockwell, Oskin, Hudnut,
  Mueller, Spelz, Akciz, Masana, Faneros et~al.}]{fletcher2014}
Fletcher JM, Teran OJ, Rockwell TK, Oskin ME, Hudnut KW, Mueller KJ, Spelz RM,
  Akciz SO, Masana E, Faneros G, et~al. (2014) Assembly of a large earthquake
  from a complex fault system: Surface rupture kinematics of {the 4 April 2010
  El Mayor--Cucapah (Mexico) Mw 7.2 earthquake}. Geosphere 10(4):797--827,
  \doi{10.1130/GES00933.1}

\bibitem[{Gao et~al.(2019)Gao, Rougier, Guyer, Lei, and Johnson}]{gao2019}
Gao K, Rougier E, Guyer RA, Lei Z, Johnson PA (2019) Simulation of crack
  induced nonlinear elasticity using the combined finite-discrete element
  method. Ultrasonics 98:51 -- 61,
  \doi{https://doi.org/10.1016/j.ultras.2019.06.003}

\bibitem[{Harris et~al.(1991)Harris, Archuleta, and Day}]{harris1991}
Harris RA, Archuleta RJ, Day SM (1991) Fault steps and the dynamic rupture
  process: {2-D} numerical simulations of a spontaneously propagating shear
  fracture. Geophys Res Lett 18(5):893--896, \doi{10.1029/91GL01061}

\bibitem[{Ida(1972)}]{ida1972a}
Ida Y (1972) Cohesive force across tip of a longitudinal-shear crack and
  griffiths specific surface-energy. J Geophys Res 77:3796--3805,
  \doi{JB077i020p03796}

\bibitem[{Knight et~al.(2015)Knight, Rougier, and Lei}]{hoss2015}
Knight EE, Rougier E, Lei Z (2015) Hybrid optimization software suite {(HOSS)}
  -- educational version {LA-UR-15-27013}. Tech. rep., Los Alamos National
  Laboratory

\bibitem[{Lachenbruch(1980)}]{lachenbruch1980a}
Lachenbruch AH (1980) Frictional heating, fluid pressure, and the resistance to
  fault motion. J Geophys Res 85(B11):6097--6112, \doi{10.1029/JB085iB11p06097}

\bibitem[{Lei and Ke(2018)}]{lei2018}
Lei Q, Ke G (2018) Correlation between fracture network properties and stress
  variability in geological media. Geophys Res Lett 45(9):3994--4006,
  \doi{10.1002/2018GL077548}

\bibitem[{Lei et~al.(2014)Lei, Rougier, Knight, and Munjiza}]{lei2014}
Lei Z, Rougier E, Knight E, Munjiza A (2014) A framework for grand scale
  parallelization of the combined finite discrete element method in 2d. Comp
  Part Mech 1(3):307--319, \doi{10.1007/s40571-014-0026-3}

\bibitem[{Lei et~al.(2019)Lei, Rougier, Munjiza, Viswanathan, and
  Knight}]{lei2019}
Lei Z, Rougier E, Munjiza A, Viswanathan H, Knight EE (2019) Simulation of
  discrete cracks driven by nearly incompressible fluid via 2d combined
  finite-discrete element method. International Journal for Numerical and
  Analytical Methods in Geomechanics 43(9):1724--1743, \doi{10.1002/nag.2929}

\bibitem[{Lisjak et~al.(2014)Lisjak, Grasselli, and Vietor}]{lisjak2014}
Lisjak A, Grasselli G, Vietor T (2014) Continuum--discontinuum analysis of
  failure mechanisms around unsupported circular excavations in anisotropic
  clay shales. Int J Rock Mech Min Sci 65:96--115,
  \doi{10.1016/j.ijrmms.2013.10.006}

\bibitem[{Mahabadi et~al.(2014)Mahabadi, Tatone, and Grasselli}]{mahabadi2014}
Mahabadi O, Tatone B, Grasselli G (2014) Influence of microscale heterogeneity
  and microstructure on the tensile behavior of crystalline rocks. J Geophys
  Res 119(7):5324--5341, \doi{10.1002/2014JB011064}

\bibitem[{Mitchell and Faulkner(2009)}]{mitchell2009}
Mitchell TM, Faulkner DR (2009) {The nature and origin of off-fault damage
  surrounding strike-slip fault zones with a wide range of displacements: a
  field study from the Atacama fault system, northern Chile}. J Struct Geol
  31(8):802--816, \doi{10.1016/j.jsg.2009.05.002}

\bibitem[{Mitchell and Faulkner(2012)}]{mitchell2012a}
Mitchell TM, Faulkner DR (2012) Towards quantifying the matrix permeability of
  fault damage zones in low porosity rocks. Earth Planet Sc Lett
  339--340:24--31, \doi{10.1016/j.epsl.2012.05.014}

\bibitem[{Munjiza et~al.(1995)Munjiza, Owen, and Bicanic}]{munjiza1995}
Munjiza A, Owen D, Bicanic N (1995) A combined finite-discrete element method
  in transient dynamics of fracturing solids. Eng Computation 12(2):145--174,
  \doi{10.1108/02644409510799532}

\bibitem[{Munjiza et~al.(1999)Munjiza, Andrews, and White}]{munjiza1999}
Munjiza A, Andrews KRF, White JK (1999) Combined single and smeared crack model
  in combined finite-discrete element analysis. Int J Numer Meth Eng
  44(1):41--57,
  \doi{10.1002/(SICI)1097-0207(19990110)44:1<41::AID-NME487>3.0.CO;2-A},
  \urlprefix\url{http://dx.doi.org/10.1002/(SICI)1097-0207(19990110)44:1<41::AID-NME487>3.0.CO;2-A}

\bibitem[{Munjiza et~al.(2015)Munjiza, Knight, and Rougier}]{munjiza2015}
Munjiza A, Knight EE, Rougier E (2015) Large strain finite element method: a
  practical course. John Wiley \& Sons

\bibitem[{Munjiza(2004)}]{munjiza2004}
Munjiza AA (2004) The combined finite-discrete element method. John Wiley \&
  Sons, \doi{10.1002/0470020180}

\bibitem[{Munjiza et~al.(2011)Munjiza, Knight, and Rougier}]{munjiza2011}
Munjiza AA, Knight EE, Rougier E (2011) Computational mechanics of discontinua.
  John Wiley \& Sons, \doi{10.1002/9781119971160}

\bibitem[{Okubo(2018)}]{okubo2018h}
Okubo K (2018) Dynamic earthquake ruptures on multiscale fault and fracture
  networks. PhD thesis, Institut de Physique du Globe de Paris

\bibitem[{Okubo et~al.(2019)Okubo, Bhat, Rougier, Marty, Schubnel, Lei, Knight,
  and Klinger}]{okubo2019}
Okubo K, Bhat HS, Rougier E, Marty S, Schubnel A, Lei Z, Knight EE, Klinger Y
  (2019) Dynamics, radiation and overall energy budget of earthquake rupture
  with coseismic off-fault damage. J Geophys Res 124, \doi{10.
  1029/2019JB017304}

\bibitem[{Palmer and Rice(1973)}]{palmer1973}
Palmer AC, Rice JR (1973) Growth of slip surfaces in progressive failure of
  over-consolidated clay. Proc R Soc Lond Ser-A 332:527--548,
  \doi{10.1098/rspa.1973.0040}

\bibitem[{Passel{\`e}gue et~al.(2016)Passel{\`e}gue, Schubnel, Nielsen, Bhat,
  Deldicque, and Madariaga}]{passelegue2016b}
Passel{\`e}gue FX, Schubnel A, Nielsen S, Bhat HS, Deldicque D, Madariaga R
  (2016) Dynamic rupture processes inferred from laboratory microearthquakes. J
  Geophys Res 121, \doi{10.1002/2015JB012694}

\bibitem[{Poliakov et~al.(2002)Poliakov, Dmowska, and Rice}]{poliakov2002}
Poliakov ANB, Dmowska R, Rice JR (2002) Dynamic shear rupture interactions with
  fault bends and off-axis secondary faulting. J Geophys Res 107(B11),
  \doi{10.1029/2001JB000572}

\bibitem[{Rice et~al.(2005)Rice, Sammis, and Parsons}]{rice2005}
Rice JR, Sammis CG, Parsons R (2005) Off-fault secondary failure induced by a
  dynamic slip pulse. Bull Seism Soc Am 95(1):109--134,
  \doi{10.1785/0120030166}

\bibitem[{Rougier et~al.(2011)Rougier, Knight, Munjiza, Sussman, Broome, Swift,
  and Bradley}]{rougier2011}
Rougier E, Knight EE, Munjiza A, Sussman AJ, Broome ST, Swift RP, Bradley CR
  (2011) The combined finite-discrete element method applied to the study of
  rock fracturing behavior in {3D}. In: 45th US Rock Mechanics/Geomechanics
  Symposium, American Rock Mechanics Association

\bibitem[{Rougier et~al.(2014)Rougier, Knight, Broome, Sussman, and
  Munjiza}]{rougier2014}
Rougier E, Knight EE, Broome ST, Sussman AJ, Munjiza A (2014) Validation of a
  three-dimensional finite-discrete element method using experimental results
  of the split hopkinson pressure bar test. Int J Rock Mech Min Sci
  70:101--108, \doi{10.1016/j.ijrmms.2014.03.011}

\bibitem[{Shipton and Cowie(2001)}]{shipton2001}
Shipton ZK, Cowie PA (2001) Damage zone and slip-surface evolution over $\mu$m
  to km scales in high-porosity navajo sandstone, utah. J Struct Geol
  23(12):1825--1844, \doi{10.1016/S0191-8141(01)00035-9}

\bibitem[{Sibson(1977)}]{sibson1977}
Sibson RH (1977) Fault rocks and fault mechanisms. J Geol Soc (London, UK)
  133(3):191--213, \doi{10.1144/gsjgs.133.3.0191}

\bibitem[{Sowers et~al.(1994)Sowers, Unruh, Lettis, and Rubin}]{sowers1994}
Sowers JM, Unruh JR, Lettis WR, Rubin TD (1994) Relationship of {the Kickapoo
  fault to the Johnson Valley and Homestead Valley faults, San Bernardino
  county, California}. Bull Seism Soc Am 84(3):528--536

\bibitem[{Templeton and Rice(2008)}]{templeton2008}
Templeton EL, Rice JR (2008) Off-fault plasticity and earthquake rupture
  dynamics: 1. dry materials or neglect of fluid pressure changes. J Geophys
  Res 113(B09306), \doi{10.1029/2007JB005529}

\bibitem[{Templeton et~al.(2009)Templeton, Baudet, Bhat, Dmowska, Rice,
  Rosakis, and Rousseau}]{templeton2009}
Templeton EL, Baudet A, Bhat HS, Dmowska R, Rice JR, Rosakis AJ, Rousseau CE
  (2009) Finite element simulations of dynamic shear rupture experiments and
  dynamic path selection along kinked and branched faults. J Geophys Res
  B08304, \doi{10.1029/2008JB006174}

\bibitem[{Thomas and Bhat(2018)}]{thomas2018a}
Thomas MY, Bhat HS (2018) Dynamic evolution of off-fault medium during an
  earthquake: a micromechanics based model. Geophys J Int 214(2):1267--1280,
  \doi{10.1093/gji/ggy129}

\bibitem[{Viesca and Garagash(2015)}]{viesca2015}
Viesca RC, Garagash DI (2015) Ubiquitous weakening of faults due to thermal
  pressurization. Nature Geosci 8(11):875--879, \doi{10.1038/ngeo2554}

\bibitem[{Viesca et~al.(2008)Viesca, Templeton, and Rice}]{viesca2008}
Viesca RC, Templeton EL, Rice JR (2008) Off-fault plasticity and earthquake
  rupture dynamics: 2. case of saturated off-fault materials. J Geophys Res
  113(B09307), \doi{10.1029/2007JB005530}

\bibitem[{Xia et~al.(2004)Xia, Rosakis, and Kanamori}]{xia2004}
Xia KW, Rosakis AJ, Kanamori H (2004) Laboratory earthquakes: The
  sub-rayleigh-to-supershear rupture transition. Science 303:1859--1861,
  \doi{10.1007/s10704-006-0030-6}

\bibitem[{Xu et~al.(2012)Xu, Ben-Zion, and Ampuero}]{xu2012a}
Xu S, Ben-Zion Y, Ampuero JP (2012) {Properties of inelastic yielding zones
  generated by in-plane dynamic ruptures---I. Model description and basic
  results}. Geophys J Int 191(3):1325--1342,
  \doi{10.1111/j.1365-246X.2012.05679.x}

\bibitem[{Zhao et~al.(2014)Zhao, Lisjak, Mahabadi, Liu, and
  Grasselli}]{zhao2014}
Zhao Q, Lisjak A, Mahabadi O, Liu Q, Grasselli G (2014) Numerical simulation of
  hydraulic fracturing and associated microseismicity using finite-discrete
  element method. J Rock Mech Geotech Eng 6(6):574--581,
  \doi{10.1016/j.jrmge.2014.10.003}

\end{thebibliography}
\end{document}